\documentclass[journal=mamobx,manuscript=article]{achemso}
\usepackage{achemso}
\setkeys{acs}{usetitle = true}
\usepackage[version=3]{mhchem}
\usepackage[T1]{fontenc}
\usepackage{times,mathptmx}
\usepackage{type1cm}
\usepackage{graphicx}
\usepackage{dcolumn}

\usepackage{bm}
\usepackage{units}
\usepackage{color}
\usepackage{xcolor}
\usepackage{latexsym}
\usepackage{subfig}
\usepackage{amsmath,amssymb}
\usepackage{caption}
\usepackage{subcaption}
\SectionNumbersOn
\setcitestyle{super}

\title{On the structure-viscoelasticity relationship of a dually crosslinked reversible polymer network}

\author{Mounika Gosika}
\affiliation{Department of Physics, School of Advanced Sciences, Vellore Institute of Technology, Vellore, 632014, Tamil Nadu, India}
\alsoaffiliation{Centro de F\'isica de Materiales (CSIC, UPV/EHU) and Materials Physics Center MPC, Paseo Manuel de Lardizabal 5, 20018 San Sebasti\'an, Spain}
\email{mounika.gosika@vit.ac.in}

\author{Angel J. Moreno}
\affiliation{Centro de F\'isica de Materiales (CSIC, UPV/EHU) and Materials Physics Center MPC, Paseo Manuel de Lardizabal 5, 20018 San Sebasti\'an, Spain}
\alsoaffiliation{Donostia International Physics Center, Paseo Manuel de Lardizabal 4, 20018 San Sebasti\'an, Spain}
\email{angeljose.moreno@ehu.eus}

\begin{document}

\newpage

\begin{abstract}
We perform equilibrium Langevin dynamics simulations to understand the relationship between 
structure and viscoelasticity of a dually crosslinked reversible polymer network. 
The dual dynamic cross-linking is achieved by introducing orthogonal crosslinkers (A and B) 
in to the polymer backbone, where only intra-species bonds (A-A or B-B) are allowed to be 
formed. We simulate the systems at infinite dilution and above the overlap concentration of 
the chains, where a system spanning network is known to be formed. Consistent with the 
previous reports, we find the dually cross-linked chains to be more compact compared to the 
singly cross-linked chains at infinite dilution. Further, at the finite concentration, we 
explore the role of weak (A-A) \textit{vs.} strong (B-B) bonds in tuning the stress relaxation behavior 
of the networks, by systematically varying the relative composition of A monomers ($x$) while 
keeping the total fraction of crosslinks as constant. Interestingly, we find a non-monotonic 
trend in the diffusivity and hence in the stress auto-correlation function of the system w.r.t. 
$x$. Furthermore, we find that in the plateau regime the dynamics is dictated by the strength 
of the weak bonds, whereas the terminal relaxation behavior of the stress auto-correlation 
function depends on the strength of the strong bonds. We also study the influence of the 
distribution of the crosslinking monomers on the stress relaxation by carrying out the 
simulations for random as well as symmetric (di-block) sequences of the chains. We observe 
that in the symmetric distribution, the intra-molecular bonds are higher than those found in 
the random distribution, due to the closer proximity of the cross-linking monomers. As a 
consequence we find the inter-molecular bonds to be significantly lower in the symmetric case 
which makes it to have a faster stress relaxation compared to its random counterpart. We 
believe our results aid in understanding the interplay of weak and strong bonds and the 
specific effects of their relative composition and distribution in tuning the viscoelasticity 
of the reversible polymer networks. 
\end{abstract}

\maketitle

\newpage

\section{Introduction}
Polymer networks embodying reversible interactions either through non-covalent or dynamic covalent bonds
among specific functional groups are ubiquitous in living systems. \cite{montero2017bioinspired, yang2018leaf, furthauer2019self} 
In addition, reversible polymer networks (RPNs) have widespread industrial applications such as self-healing 
materials and recyclable plastics, owing to the re-processability offered by the cross-linking units. \cite{ma2021ultra,jin2019malleable}
Of specific interest are the RPNs based on dynamic covalent bonding, such as 
\textit{vitrimers}\cite{montarnal2011silica,zheng2021vitrimers}, which have properties intermediate to thermosets 
(mechanically robust but cannot be reshaped) and thermoplastics (malleable but not stable when exposed to solvents 
or at high temperatures). Recently, single-chain nanoparticles (SCNPs) are becoming popular in the class of 
RPNs \cite{verde2020single,formanek2021gel,rovigatti2022designing}. Generally, the reversible or irreversible 
crosslinks between the functional groups within the SCNPs can be triggered with the aid of pH, temperature, 
catalysts, or non-bonding interactions such as hydrogen-bonding and metal-ligand coordination, \cite{lyon2015brief,verde2020single}, 
at dilute concentrations. The typical sizes of the experimentally synthesized SCNPs range from 3 to 30 nm and 
the SCNPs are gaining attention in drug delivery, catalysis, biosensors, and as rheology modifying agents 
among others \cite{nitti2022single}. 

The structure and dynamics of reversible SCNPs in solutions have been investigated using small-angle neutron 
scattering experiments as well as molecular dynamics simulations \cite{arbe2016structure, gonzalez2020structure, moreno2017computer}. The studies reveal that the standard synthesis routes used for the cross-linking such as 
Michael addition \cite{sanchez2013michael}, copper complexation \cite{sanchez2014metallo}, generate sparse SCNPs at 
dilute concentrations. This is a consequence of the formation of self-avoiding conformations of the SCNPs under good 
solvent conditions. However, for catalysis, bio-medicine applications, and for enzyme modeling, the globular morphology 
of the SCNPs is preferred. \cite{pomposo2014far, hamelmann2023single} Introducing orthogonal cross-linkers in the 
SCNP's backbone is one such approach to circumvent the problem, as the formation of the long-ranged loops among the 
functional groups, a key aspect for the compaction of the SCNP, is favored with such a cross-linking 
\cite{moreno2013advantages, blazquez2021advances}. A recent simulation study\cite{formanek2021gel} on these reversible SCNPs
based on a single reactive monomer cross-linking investigated the structure of the SCNPs ranging from a high dilution 
limit to densities well above the overlap concentration of the SCNPs. The key observation from the study is that the 
molecular conformations of these reversible SCNPs are weakly perturbed under crowding, as the structure retains the sparse
conformations through the formation of inter-molecular bonds at high concentrations. Therefore, the "effective-fluid" 
approach, where the interactions between the "fluid" particles are described using the effective pair 
interaction potentials of the SCNPs, was shown to hold well at densities far above the overlap concentrations of the SCNPs \cite{paciolla2022validity}. 
In another computational work Rovigatti et al \cite{rovigatti2022designing} investigated the molecular conformations of 
the multi-orthogonally cross-linked SCNPs. The authors showed that the conformations of the SCNPs become more compact when the 
type of cross-linkable units is increased along the SCNP backbone. Further, the effective interactions between two 
SCNPs involving distinct types of reactive monomers are shown to be attractive in nature, whereas those between the SCNPs with a single type of reactive monomer are repulsive in nature. This study also demonstrates that liquid-gas phase separation is possible in an SCNP melt when the SCNP is designed to have multi-orthogonal cross-linkers in its backbone. \cite{rovigatti2022designing} 

In the context of multi-orthogonal cross-linking, the reversible bond energy of a given type of reactive monomer pair is modeled to be the same in all the above-mentioned studies. \cite{rovigatti2022designing, moreno2013advantages} Moreover, as RPNs have mechanical properties intermediate to that of thermosets and thermoplastics, a thorough investigation of the factors which have the 
potential to tune to the viscoelastic behavior of the RPNs is quintessential, alongside the structural aspects.  To this end, Ciarella et al \cite{ciarella2018dynamics} 
reported that stress relaxes faster in a star-polymer network that has loop defects (two ends of the star are bound together) 
in its architecture, highlighting the role of the network topology on stress relaxation. Traditionally in elastomers, a fraction 
of \textit{sacrificial bonds} (bonds that break first under mechanical stress) were introduced to improve the toughness of 
the networks \cite{ducrot2014toughening}. Recently, in the same context, double dynamic polymer networks (DDPN) are becoming popular \cite{cui2022double}. The DDPNs have two distinct dynamics and can be formed in two ways: either through i) two interpenetrated networks or by ii) a single network consisting of orthogonal cross-linking monomers. By modulating the factors which control the microscopic dynamics of fast \textit{vs.} slow bonds in these DDPNs the macroscopic viscoelastic response can be tuned. \cite{van2022preface} 

For instance, in an experimental work, Chen et al \cite{chen2019rapid} synthesized a dual dynamic 
epoxy vitrimer, wherein simultaneous exchange of carboxylate ester bonds and disulfide bonds can occur when the temperature is used
as a trigger. The authors reported that the stress relaxes faster in the double vitrimer network as compared to the single networks 
of disulfide and ester bonds.  Similarly, in dual hydrogel networks consisting of a loosely crosslinked covalent network and a dense network of transient H-bonds, the dissociation rate of the transient bonds is shown 
to control the rigidity of the networks. \cite{hu2017dynamics, zhao2022fracture} 
Here, the faster dissociation of the transient bonds over the permanent ones aids in dissipating the energy at high strain rates 
and improves the toughness of the material. \cite{cao2020mechanical, cui2022double} Specifically, the fraction, lifetime, and bond strength of these transient
bonds are reported as the key parameters in tuning the stress relaxation of the DDPNs formed by combining permanent and reversible bond interactions. \cite{cao2020mechanical, raffaelli2021stress}  Therefore, synergistic incorporation of two or more types of reversible bonding schemes in DDPNs can modify their mechanical properties dramatically  \cite{araya2014reversible, lai2018rigid, delpierre2019simple, hammer2021dually}.

Furthermore, in DDPNs comprising of stronger permanent bonds and relatively weaker transient bonds, the weaker bonds allow for a rearrangement of the network, while the stronger bonds help in stabilizing the networks, thus catering to both the self-healing and mechanical strength aspects of the networks. \cite{zhang2022progress}  
As mentioned above, the central idea of the works based on utilizing two or more reversible interactions to improve 
the self-healing properties of the materials lies in taking advantage of the interplay of the weak \textit{vs.} strong bonds 
in the network. A systematic investigation of how the relative composition of weak \textit{vs.} strong bonds influence the stress
relaxation behavior, alongside their strength is lacking. To this end, in this article, we perform an in-silico investigation of the structural and dynamical aspects of a model SCNP RPN,
with dual crosslinkers (A and B), where only A-A and B-B type reversible bonds are allowed to be formed. Moreover, to model the weak
\textit{vs.} strong bonds, we deliberately make the reversible bond energy of A monomers (weak) to be always less than that of the 
B monomers (strong). Specifically, we address the influence of the reversible bond energy disparity, relative composition, 
and distribution of the weak \textit{vs.} strong bonds on the structure of the SCNPs at infinite dilution and at a concentration 
above the overlap concentration of the polymers, using equilibrium Langevin dynamics simulations. Further, at the finite 
concentration we also discuss how the structure and bonding dynamics of the networks influences their viscoelastic response.  

The remainder of the article is organized as follows. In section 2, we provide 
the details of the polymer model and the simulation protocol implemented, along with the details of the systems studied. 
In section 3, we discuss our simulation results on the size, bonding probabilities, and the intra \textit{vs.} inter-molecular bonding 
competitions of the dually cross-linked SCNPs as a function of the reactive monomer distribution 
as well as the bond energy disparity between weak and strong bonds. We also describe the dynamic properties such as the inter-bond correlation times and its influence on the viscoelastic response of the networks, quantified through stress
auto-correlation function. Finally, in section 4, we summarize the major findings from our study and 
provide future directions. 

\section{Simulation Details}
\subsection{Polymer Model and Interaction Potentials}
We have used the well-known Kremer-Grest model for the polymer chains 
\cite{kremer1990dynamics}. The system consists of $N_p=108$ polymer chains, each of which has 
$N_m=200$ monomers. A fraction $f=0.3$ of the $N_m$ monomers are reactive sites. If two reactive 
sites are within a cut-off distance ($r_{\rm c} < 1.3$), they form a reversible bond. The 
details of the model and the interaction potentials are thoroughly discussed in previous 
works \cite{formanek2021gel,paciolla2022validity}. We briefly mention the main details here. 
All the monomers within a given polymer chain are irreversibly bonded to their adjacent 
monomers through the finitely extensible non-linear elastic (FENE) potential defined as:
\begin{equation}
V^{\rm FENE}(r) = -\epsilon K_{\rm F} R_0^2 \ln \left[ 1 - \left(\frac{r}{R_0}\right)^2\right],
\end{equation}
where $K_{\rm F} =15$ and $R_0 =1.5$. The excluded volume interactions between both bonded and non-bonded monomers are modeled via the Weeks-Chandler-Andersen potential defined as:
\begin{equation}
V^{\rm WCA}(r) = 
	\begin{cases}
	4\epsilon\left[\left(\frac{\sigma}{r}\right)^{12} - \left(\frac{\sigma}{r}\right)^6 + \frac{1}{4} \right] & \text{for} \,\, r \le 2^{1/6}\sigma \\
        0   & \text{for} \,\, r > 2^{1/6}\sigma ,
	\end{cases}
\end{equation}
which is purely repulsive due the chosen cut-off at $r = 2^{1/6}\sigma$. The quantities
$\epsilon$ and $\sigma$ are the units of the energy and distance, respectively. In the rest 
of the article, we will use $\epsilon= \sigma =1$. 

If two reactive monomers are forming a reversible bond they mutually interact through 
a Stillinger-Weber type of potential\cite{rovigatti2018self}, defined as:
\begin{equation}
V_{\rm bond}(r) = 
\begin{cases}
	\epsilon_{\rm b} \, e^{\sigma/(r_{\rm c} - \sigma)} \left[A\left(\left({\frac{\sigma}{r}}\right)^4 -1\right) - 1\right] \, e^{\sigma/(r - r_{\rm c})} & \text{for} \,\, r \le r_{\rm c} \\
	0 & \text{for} \,\, r > r_{\rm c} \label{eqrevb}
\end{cases}
\end{equation}
where $r_{\rm c} =1.3 \sigma$ is the capture radius and $A=\frac{\sigma^2}{4(r_{\rm c} - \sigma)^2}$. This potential is short-ranged and has a minimum at $r_{\rm min} = \sigma$

In order to limit the valence to a single reversible bond per reactive site, we implement an 
additional 3-body potential defined as \cite{sciortino2017three}: 
\begin{equation}
V_{\rm 3b}(r) =  \sum_{ijk} (\epsilon_{{\rm b},ij} \epsilon_{{\rm b},ik})^{1/2} \Phi_3(r_{ij}) \Phi_3(r_{ik}). \label{v3b}
\end{equation}
This potential is switched on when a reactive site $k$ enters the capture radius of a 
reactive site $i$ that is already bonded to another one $j$. The function $\Phi_3$ is 
defined as:
\begin{equation}
\Phi_3(r_{ij}) = 
	\begin{cases}
	1  & \text{for} \, \, r < r_{\rm min} \\
	-V_{\rm bond,ij}(r)/ \epsilon_{\rm b,ij} & \text{for} \, \, r_{\rm min} \le r \le r_{\rm c}\\
	0  & \text{for} \, \, r > r_{\rm c}. \\
	\end{cases}
\end{equation} 
By construction, when a triplet is formed the energy decrease associated to the new bond is 
compensated by the 3-body repulsive term without changing the potential energy of the system. 
As a consequence, the triplets are very short-lived and bond swapping is facilitated,
speeding up the exploration of different patterns of the bond network. Moreover, monovalent 
bonding is governed by a Hamiltonian, unlike methods based on random choices when one site 
can bind to more than one candidate.

\begin{figure}
	\centering
	\includegraphics[scale=0.5]{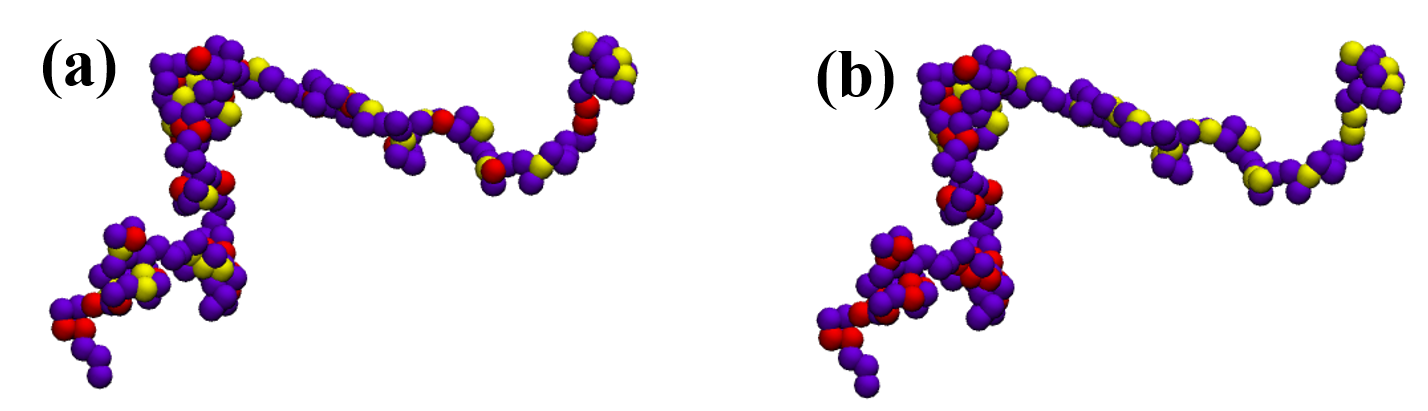}
	\caption{Typical initial snapshots of an isolated polymer chain with $N_{\rm r}=60$ reactive sites,
	 with $N_{\rm A}=N_{\rm B}=30$.
	The reactive A and B-monomers are depicted in red and yellow colors, respectively. The purple-colored monomers are non-reactive. Panels (a) and (b) correspond to the random and symmetric distributions, respectively.}
	\label{figsyrand}
\end{figure}

The details of the simulated systems are given in Table~\ref{tab:1}. To understand the 
effects of the dual dynamic bond exchange we label the reactive monomers as A and B. By 
construction, only A-A and B-B type reversible bonds are allowed to be formed. We investigate 
the influence of the weak and strong bonds on the structural and dynamical properties of the 
system by assigning two different energies to the A-A and B-B reversible bonds. In other 
words, we analyze the effect of the bond energy disparity by selecting two values of the pair 
$(\epsilon_{\rm b,AA}, \epsilon_{\rm b,BB})$, namely $(\epsilon_{\rm b,AA}, \epsilon_{\rm b,BB}) = (12,15)$ and (10,17).
Furthermore, we analyze the effect of the composition by investigating, at a fixed fraction 
of reactive monomers, three different fractions of the weak bond-former A-monomers, $x= N_{\rm A}/N_{\rm r}$ with $N_{\rm A}$ the number of A-monomers in each chain. 
Namely, we investigate the cases $x= 0.17, 0.5 $ and $0.83$. Finally, at fixed equimolar 
composition, $x=0.5$, we investigate two types of sequences of A and B-sites. In both of them 
the number of A- and B-sites is identical but, while in the first case (labeled as `random') 
A- and B-sites are distributed randomly along the whole chain contour, in the second case 
(labeled as `symmetric') they follow a 'diblock' structure, i.e., A- and B-sites are 
distributed randomly within the first and second half of the chain respectively. 
In all cases, to prevent reversible bonds between already irreversibly connected monomers, 
the sequence is constructed with the condition that two consecutive reactive monomers are 
separated by at least one non-reactive monomer of a reactive one of the other species.
Typical initial snapshots of an isolated polymer chain with $x=0.5$ corresponding to the
random and symmetric distributions are shown in Fig. \ref{figsyrand} (a) and (b), 
respectively.

\begin{table}[h!]
	\centering
	\small
	\begin{tabular}{|c|c|c|c|c|}
		\hline
		$x$ & Number of reactive monomers & $x=\frac{N_A}{N_r}$ & Reversible bond potential  & Monomer distribution \\
		& A, B &  & parameters: $\epsilon_{b,AA}$ and $\epsilon_{b,BB}$  &  \\ 
		&     &   &  (units of $\epsilon$) & \\ \hline
		0.17  & 10, 50 & 0.17 & 12 and 15 & random \\
		&    & & 10 and 17 & random \\ \hline 
		0.50 & 30, 30 & 0.5 & 12 and 15 & random and symmetric \\ 
		&    & & 10 and 17 & random \\ \hline 
		0.83 & 50, 10 & 0.83 & 12 and 15 & random \\
		&    & & 10 and 17 & random \\ \hline
	\end{tabular}
	\caption{List of the systems studied. $x$ is the relative composition of A monomers i.e., $\frac{N_A}{N_r}$, 
	where $N_A$ is the number of $A$ monomers and $N_r$ is the total number of reactive monomers. $N_r=N_A + N_B=60.$}
	\label{tab:1}
\end{table}

\subsection{Simulation Methodology}

The 108 polymer chains at the different compositions and bond energies mentioned in Table \ref{tab:1}, are 
initially placed randomly in a large simulation box, avoiding overlaps 
(within a tolerance of $0.9 \sigma$). The systems are then sequentially compressed to the 
target densities, after which the Langevin dynamics simulations in an NVT ensemble are 
performed. The temperature was maintained at $T=\epsilon/k_B=1.0$ using a 
Langevin thermostat of a friction coefficient $\gamma=0.05$ \cite{smith2009dl}. The equations 
of motion are integrated using a scheme of Ref.\cite{izaguirre2001langevin} with a timestep 
of the simulation, $\delta t=0.01$. The box size is chosen as $L_{box}=50$. The corresponding 
density ($\rho=\frac{N_{p} \times N_{m}}{L_{box}^3}=0.173$) is well above the overlap 
concentration of these SCNPs ($\rho^{\ast}=N_{m}/(2\,R_{g0,\text{pure A}})^3=0.03$) and hence 
a system spanning network is known to form \cite{formanek2021gel}. For each of the systems in 
Table \ref{tab:1}, we have performed the simulations of 8 independent initial configurations 
of the system. Typical lengths of the equilibration and production runs are $10^7$ and 
$5 \times 10^7$ steps, respectively. At $x=0.5$, we also studied the density dependence on 
the size and bonding probabilities of the chains by performing additional simulations 
for $L_{box}=55, 64, 77, 103$ and $175$ \textit{i.e.,} at $\rho=0.13, 0.082, 0.047, 0.02$ and $0.004$, 
respectively, for both random and symmetric distributions of the reactive monomers. 

\section{Results and Discussion}
\subsection{Influence of the relative composition and bond strength}
One of the motivations of our study is to investigate the effects of weak \textit{vs.} strong 
bonds on the structure of the network. Therefore, we begin by discussing the static 
properties \textit{i.e.,} the size, bonding probabilities, and contour distance 
distributions, across the different scenarios studied (Table \ref{tab:1}) at infinite 
dilution (experimentally relevant, as the SCNPs are typically synthesized at this 
concentration) and at a finite concentration ($\rho=0.173$) which is well above the overlap concentration ($\rho^{\ast}$=0.03) of the polymers. 

\subsubsection{Size and bonding probabilities}
\textbf{(a) Isolated Chains} \newline
To mimic the system at infinite dilution, we have performed the simulations by avoiding 
inter-molecular interactions. The preliminary insight into the size of the chains 
can be obtained from the radius of gyration ($R_g$). As can be noted from Table \ref{tab:2} 
the size of the dual-dynamic or hetero cross-linked chains is relatively lesser than the 
singly reactive or homo cross-linked chains (pure A in Table \ref{tab:1}), confirming the 
observation that orthogonal folding is an effective strategy to obtain compact SCNPs \cite{moreno2013advantages,blazquez2021advances}. Moreover, we find a slight non-monotonic dependence 
on the size of the chains w.r.t $x$ for both $(\epsilon_{\rm b,AA}, \epsilon_{\rm b,BB}) = (12,15)$ 
and (10,17). At a fixed monomer composition, we 
do not observe any significant change in the size when the bond energy disparity is changed for 
these random distributions of the reactive monomer cases studied. The detailed description of the
molecular architecture of the SCNPs is provided in Section 3.1.2 (Molecular conformations and internal structure). 

To understand these trends in detail, we computed the bonding probability ($p_{\rm B}$), which 
quantifies the number of monomers involved in forming reversible bonds (Table \ref{tab:2}). 
Precisely, $p_{\rm B}$ denotes the ratio of the number of monomers involved in forming the reversible 
bonds to the total number of reactive monomers of the chain. Similarly, $p_{\rm B,A}$ and $p_{\rm B, B}$
denote the probabilities of forming A-A and B-B reversible bonds, respectively. From Table \ref{tab:2}, 
we can infer that $p_{\rm B0}$ of these isolated chains 
does not change significantly w.r.t. $x$ at any given value of $(\epsilon_{\rm b,AA}, \epsilon_{\rm b,BB})$.
The total number of bonds (intra-chain at $\rho=0$) per molecule also reflects the same trend w.r.t $x$, 
for both the bond energy disparities studied (black curves in Fig. S1). Although no significant changes
are noted in the bonding probabilities of the stronger bonds ($p_{\rm B0,B}$ in Table \ref{tab:2}), 
we find a significant reduction in the probability of weaker bond formation ($p_{\rm B0,A}$ in Table \ref{tab:2}),
when the bond-energy disparity is increased. This results in an overall reduction in the $p_{\rm B0}$ values in
$(\epsilon_{\rm b,AA}, \epsilon_{\rm b,BB})$ = (10,17) case as compared to that in (12,15) case.
All the above-mentioned trends are
also reflected in the A-A, B-B, and total number of bonds w.r.t $x$ and $(\epsilon_{\rm b,AA}, \epsilon_{\rm b, BB})$ illustrated in Fig. S1. 

\begin{table}[h!]
    \centering
    \begin{tabular}{|c|c|c|c|c|c|}
		\hline 
		$x$ &  $\epsilon_{\rm b,AA}$, $\epsilon_{\rm b,BB}$ & $R_{\rm g0}$ & $p_{\rm B0}$ & $p_{\rm B0,A}$ & $p_{\rm B0,B}$ \\ \hline 
		0.17    & 12, 15 & 7.87 $\pm$ 0.09 & 0.93 & 0.80 & 0.96 \\
		& 10, 17 & 7.96 $\pm$ 0.09 & 0.91 & 0.63 & 0.97 \\ \hline
		0.5    & 12, 15 & 7.42 $\pm$ 0.07 & 0.94 & 0.91 & 0.97 \\
		& 10, 17 & 7.50 $\pm$ 0.08 & 0.91 & 0.83 & 0.98 \\ \hline
		0.83   & 12, 15 & 7.81 $\pm$ 0.09 & 0.93 & 0.92 & 0.95 \\
		& 10, 17 & 7.77 $\pm$ 0.09 & 0.89 & 0.87 & 0.98 \\ \hline
		1 & 12 & 9.35 $\pm$ 0.15 & 0.92 & 0.92 & - \\ \hline
	\end{tabular}
	\caption{Size ($R_{\rm g0}$) and bonding probabilities ($p_{\rm B0}$) of the chains at infinite dilution for the cases studied in Table \ref{tab:1}. The subscript `0' denotes that the values correspond to the system of isolated chains. Similarly, $p_{\rm B0, A}$ and $p_{\rm B0,B}$ denote the probabilities of forming A-A and B-B bonds.}
	\label{tab:2}
\end{table}

\noindent \textbf{(b) Concentrated chains} \newline
Now, we discuss the above-mentioned static properties for the concentrated chains 
($\rho=0.173$), which are summarized in Table \ref{tab:3}. The size of the interacting chains
does not vary significantly w.r.t $x$ and the bond energy disparity. Not surprisingly, as the chains 
also participate in forming inter-molecular bonds in addition to the intra-molecular ones at this concentration \cite{formanek2021gel}, their sizes are larger than those of the isolated chains (Table \ref{tab:2}). 
Consistently, $p_{\rm B}$ is slightly greater than 
those found for the respective case of non-interacting chains (Table \ref{tab:2}), suggesting that these 
concentrated chains participate in inter-chain 
bonds in addition to the intra-chain bonds. Indeed, as can be seen in Fig. S1 (b) and (c), while the majority of the bonds
formed are intra-molecular in nature, about $23 \%$ of the reactive monomers are involved in forming inter-chain bonds. 
Other observations found for the isolated chains, that the i) $p_{\rm B}$ being relatively more for 
($\epsilon_{\rm b,AA}$, $\epsilon_{\rm b,BB}$) = (12,15) case, ii) weak dependence of $p_{\rm B}$ on the relative 
composition, iii) decrement in the probability of forming weaker bonds when the bond energy disparity is increased, and iv) $p_{\rm B, A} \le p_{\rm B, B}$ hold true for the interacting 
chains as well. Hence, we can conclude that the total bonding probability of the dual networks can be controlled by both the fraction as well as the relative strength of the weaker bonds. 

\begin{table}[h!]
	\centering
	\begin{tabular}{|c|c|c|c|c|c|}
		\hline 
		$x$ &  $\epsilon_{\rm b,AA}$, $\epsilon_{\rm b,BB}$ & $R_{\rm g}$ & $p_{\rm B}$ & $p_{\rm B,A}$ & $p_{\rm B,B}$ \\ \hline 
		0.17    & 12, 15 & 9.01 $\pm$ 0.07 & 0.95 & 0.89 & 0.96 \\
		          & 10, 17 & 8.99 $\pm$ 0.07 & 0.93 & 0.76 & 0.96 \\ \hline
		0.50    & 12, 15 & 8.91 $\pm$ 0.04 & 0.95 & 0.94 & 0.97 \\
		        & 10, 17 & 8.95 $\pm$ 0.05 & 0.92 & 0.87 & 0.98 \\ \hline
		0.83    & 12, 15 & 8.94 $\pm$ 0.04 & 0.94 & 0.93 & 0.97 \\
		        & 10, 17 & 9.01 $\pm$ 0.07 & 0.91 & 0.89 & 0.98 \\ \hline
	\end{tabular}
	\caption{Size and bonding probabilities of the chains at $\rho=0.173$ for the random distribution of the reactive monomer cases studied.}
	\label{tab:3}
\end{table}

\subsubsection{Molecular conformations and internal structure}
The nature of the molecular conformations of the SCNPs is dependent on whether the 
reactive monomers are involved in short-range (extended conformations) or long-ranged loops 
(compact conformations). Hence, to obtain a deeper understanding of the conformations of the 
chains we computed the contour distance distributions ($P(s)$). The contour distance between a 
reactive monomer pair forming a bond is defined as $s=\lvert i - j \rvert$, where $i$ and $j$ 
are the indices of the reactive monomers. After computing the contour distances, we obtain a 
histogram of them. Finally, $P(s)$ is calculated after normalizing the histogram frequencies 
with the total number of contour distances recorded. Hence, by definition $\sum_{s=1}^{N-1} 
P(s) = 1$. 

\begin{figure}[ht!]
    \centering
    \includegraphics[width=0.5\textwidth]{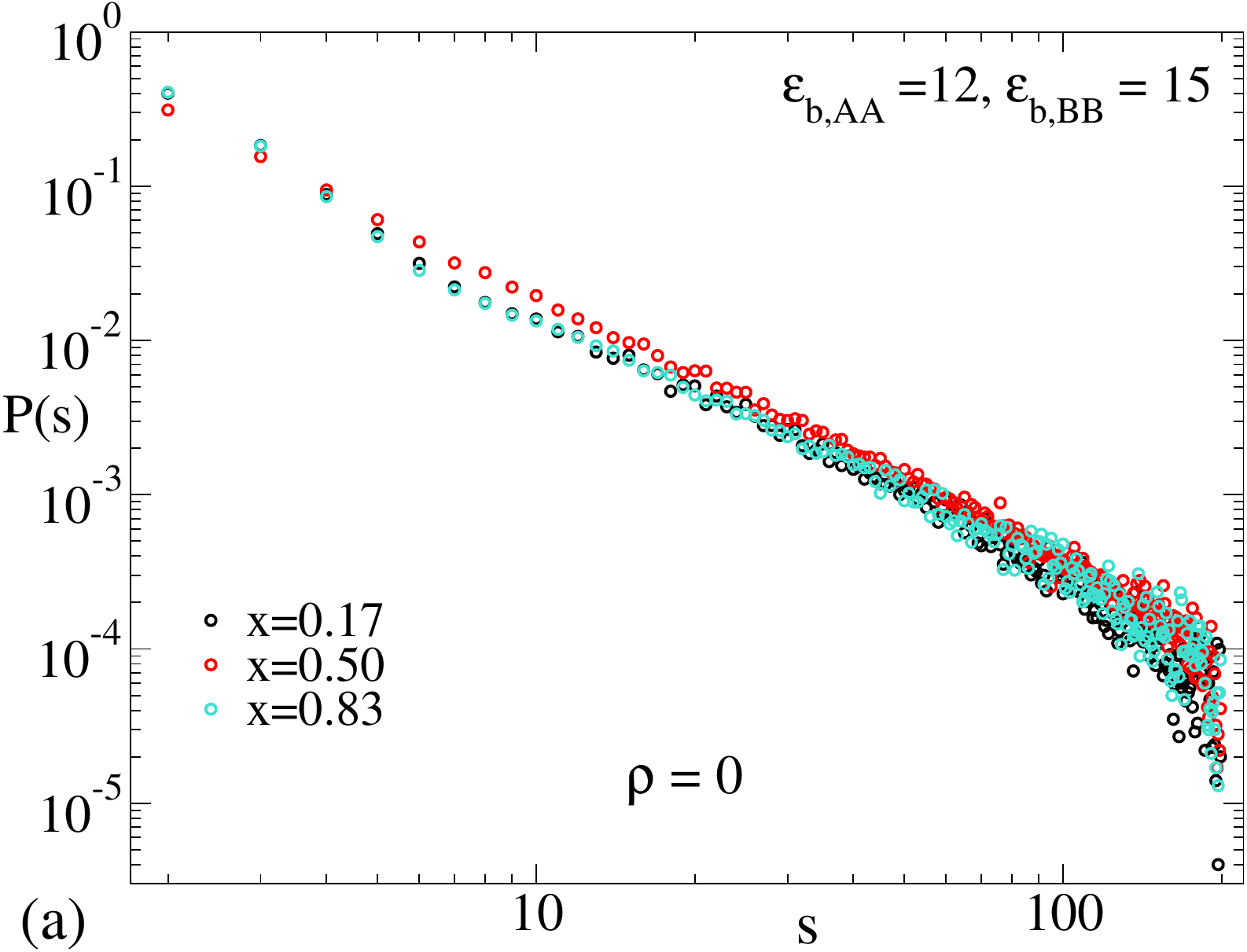}
    \\ [2 mm]
    \includegraphics[width=0.5\textwidth]{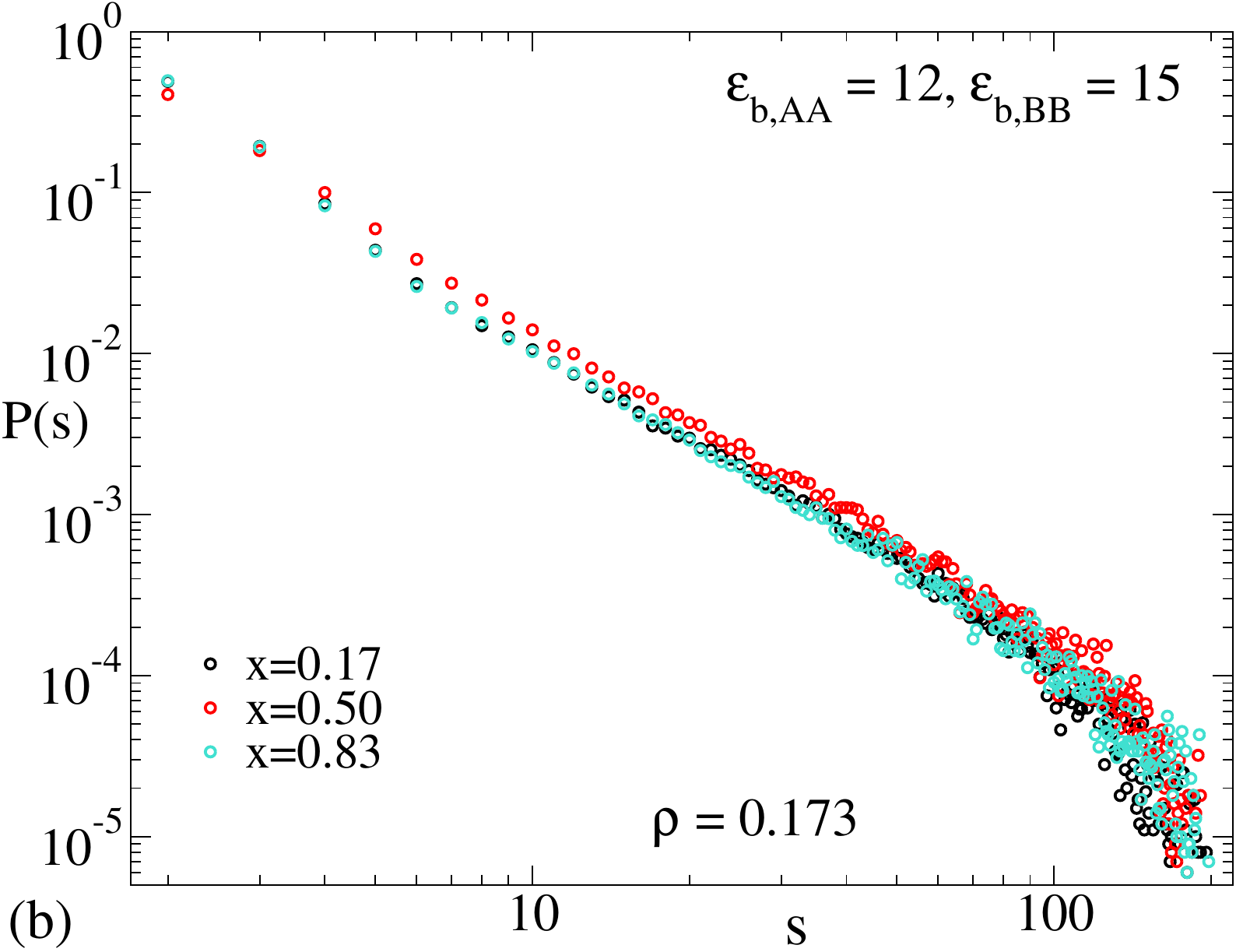}
    \\ [2 mm]
    \caption{Contour distance distributions w.r.t. $x$ at (a) $\rho = 0$ and (b) $\rho = 0.173$, for
    $(\epsilon_{\rm b,AA}, \epsilon_{\rm b,BB})=(12,15)$}
    \label{fig:cdistisoconc}
\end{figure}

In Fig.\ref{fig:cdistisoconc}, we present the $P(s)$ profiles for the isolated and concentrated chains 
for ($\epsilon_{\rm b,AA}, \epsilon_{\rm b, BB}$) = (12,15). Interestingly, in $x=0.5$ case the 
short-range loops ($1<s<4$) are relatively less and the long-ranged loops ($s>5$) are consistently large 
as compared to $x=0.17$ and $0.83$, at both the concentrations. This is because at $x=0.17$ and $x=0.83$, 
the number of majority species that can form reversible bonds are more (A=50 in $x=0.17$ and B=50 in $x=0.83$) 
and therefore involve in relatively more short-ranged loops compared to $x=0.5$ case. In $x=0.5$ case, 
due to the reduction in the number of cross-linking monomers of a given type (A=30 or B=30), long-ranged loops are 
promoted (Fig. \ref{fig:cdistisoconc}). Indeed, this is the reason for the observed non-monotonicity 
in the $R_{\rm g}$ of the chains w.r.t. $x$  and the lesser $R_g$ value at $x=0.5$ (section 3.1).
Thus, at a fixed $f$, the relative magnitude of the majority species directly impacts the compactness
of the SCNP formed, as it influences the 
formation of short-ranged \textit{vs.} long-ranged loops. As we go towards higher densities, the long-ranged 
intra-molecular bonds can be exchanged for an inter-molecular bond, owing to a gain in combinatorial entropy \cite{rovigatti2022designing}. 
As a consequence, we find that the probability of forming long-ranged loops is relatively less at $\rho=0.173$,
as compared to that at infinite dilution (Fig. \ref{fig:cdistisoconc} (a) and (b)). At the finite concentration
also we find the probability of forming long-ranged loops to be more at $x=0.5$. Further, we find the number of
inter-chain bonds (Fig. S1 (c)) to be relatively more at $x=0.5$. This finding reveals that greater the contour
distance between the alike reactive monomers, greater is the probability of forming long-ranged loops and inter-chain
bonds. 

The trends in the short-ranged \textit{vs.} long-ranged loops w.r.t $x$ are not 
dependent on the bond energy disparity, as we find similar $P(s)$ profiles like in Fig. \ref{fig:cdistisoconc} at a 
given $x$ for ($\epsilon_{\rm b,AA}, \epsilon_{\rm b, BB}$) = (10,17) case at both the concentrations studied (not shown). 

The intra-molecular structure of the polymer chains can be obtained from the form factor defined as,
\begin{equation}
    w(q) = \left \langle \frac{1}{N} \sum_{j,k} \frac{\sin(q r_{jk})}{q r_{jk}} \right \rangle.
\end{equation}
The indices $j$ and $k$ run over all the monomer pairs of a polymer chain with $N$ monomers. 
$r_{jk}$ is the distance between the $j$ and $k$ monomers. In the fractal regime one can fit 
$w(q) \sim q^{-1/\nu}$, to obtain the Flory exponent. 

\begin{figure}[ht!]
    \centering
    \includegraphics[width=0.5\textwidth]{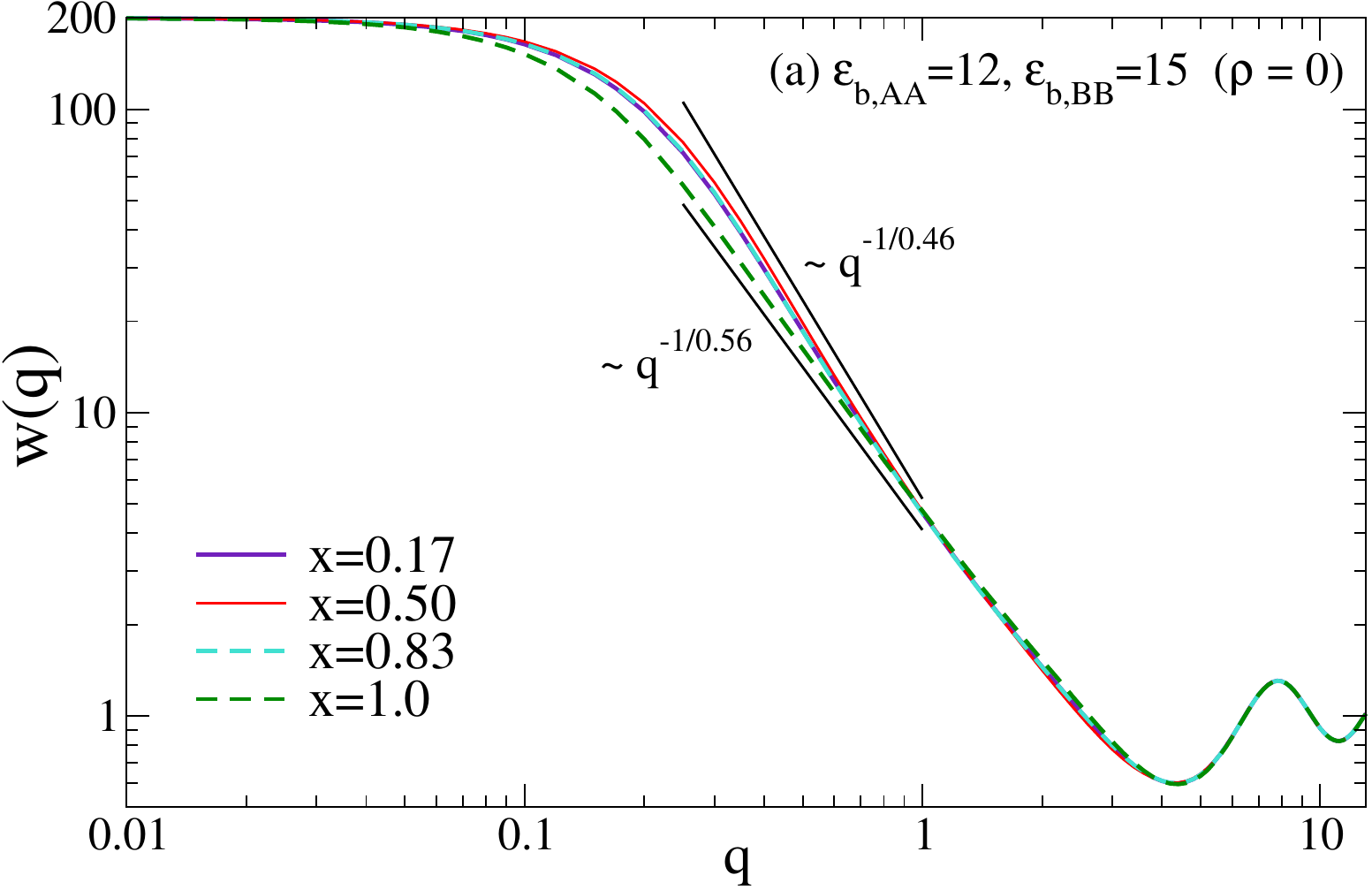}
    \\ [2mm]
    \includegraphics[width=0.5\textwidth]{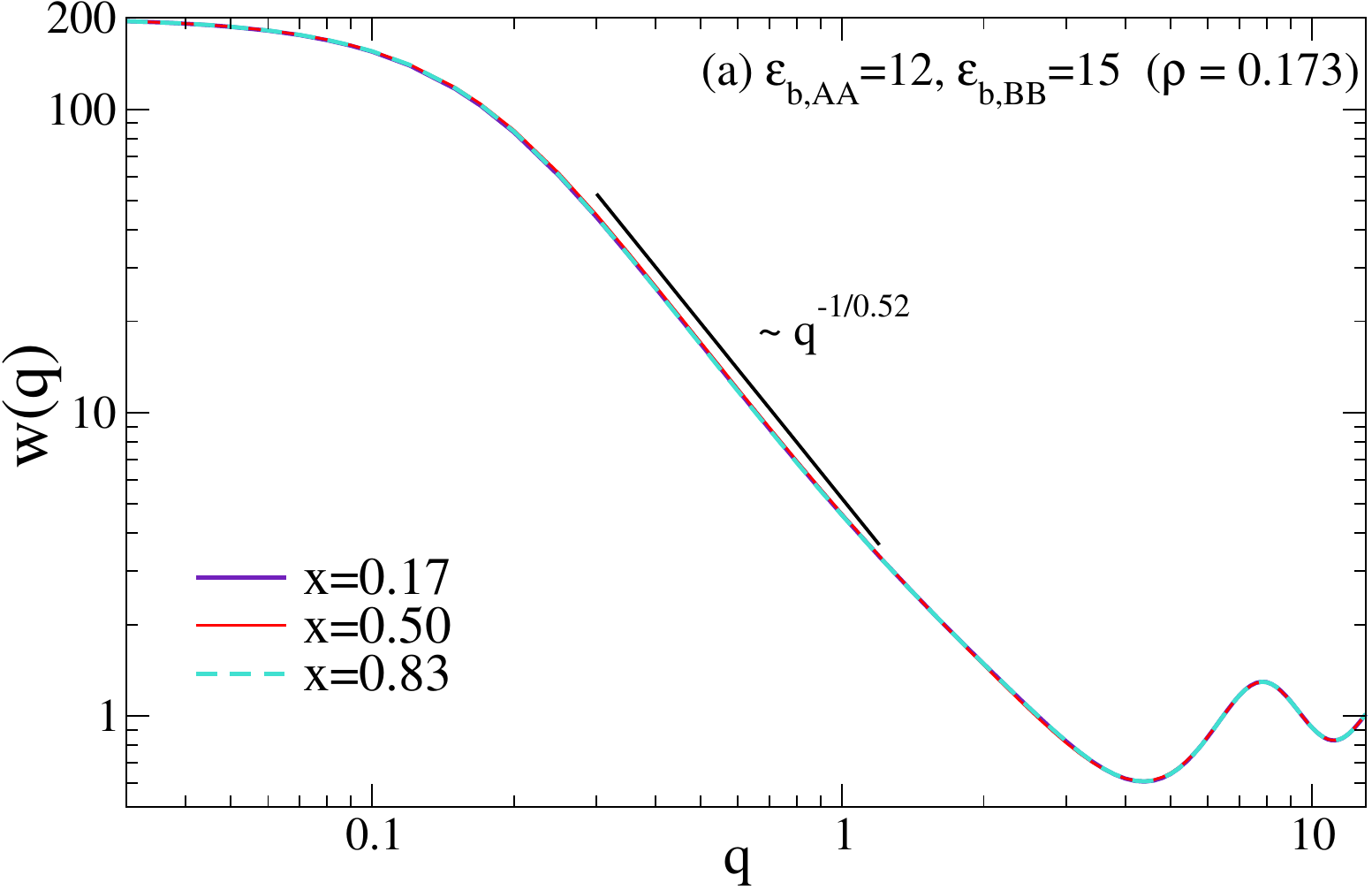}
    \caption{Form factor profiles of the individual chains of ($\epsilon_{\rm b,AA}, \epsilon_{\rm b, BB}$) = (12,15) system at (a) $\rho=0$, and (b) $\rho=0.173$.}
    \label{fig:wqisoconc}
\end{figure} 

As can be seen from Fig. \ref{fig:wqisoconc} (a), 
at the dilute concentration, the dually crosslinked chains collapse significantly, as the exponent is 
relatively lesser ($\nu \sim 0.46$), when compared with that of the singly-crosslinked case where the chains 
adopt extended conformations ($\nu \sim 0.56$). Therefore, introducing orthogonal cross-linkers is indeed an 
effective approach for generating more compact SCNPs, in agreement with the reported studies \cite{moreno2013advantages,blazquez2021advances}. 
At the finite concentration, due to the formation of inter-chain bonds (Fig. S1 (c)) and more short-ranged loops compared to the $\rho=0$ case (Fig. \ref{fig:cdistisoconc}), 
we do not observe any collapse ($\nu \sim 0.52$, Fig. \ref{fig:wqisoconc} (b)). The Flory exponent at the finite 
concentration is not significantly different as compared to that of the isolated chains. This observation is 
consistent with the previous reports where the molecular conformations of the SCNPs were found to be weakly
perturbed under crowding. \cite{formanek2021gel}
Further, we can note from Fig. \ref{fig:wqisoconc} that the internal architecture adopted by the dual-crosslinked chains 
is not dependent on the monomer composition. Similar exponents 
were found for the isolated chains ($\nu \sim 0.48$) and concentrated chains ($\nu \sim 0.52$) for ($\epsilon_{\rm b,AA}, \epsilon_{\rm b, BB}$) = (10,17) case (Fig. S2), suggesting the weak dependence of the 
bond energy disparity on the internal architecture adopted by the chains at both the concentrations studied. 

\subsection{Influence of the monomer distribution}
So far we have highlighted the effects of the monomer composition and the bond energy disparity on the structure
and the bonding probability of the SCNPs. As discussed in the introduction, the reactive monomer
distribution also significantly affects the structure and network topology. Therefore, to understand the 
influence of the monomer distribution, we performed an additional set of simulations at $x=0.5$ with a symmetric 
configuration of the chains, where the first half of the 30 reactive monomers are of type A and the rest of the 
30 reactive monomers are of type B (see Fig. \ref{figsyrand}) \textit{i.e.,} a di-block type of arrangement of the 
reactive monomers along the polymer backbone. Moreover, to understand the density dependence on the size of the 
chains, we performed the simulations for $L_{box} = 55, 64, 77, 103$, and $175$ the corresponding 
densities being $\rho=0.13, \, 0.082, \, 0.047, \, 0.02, \text{and} \,\, 0.004$, respectively, for both 
symmetric and random distributions of the reactive monomers, at $x=0.5$. 

\begin{figure}[h!]
    \centering
    \includegraphics[width=0.45\textwidth]{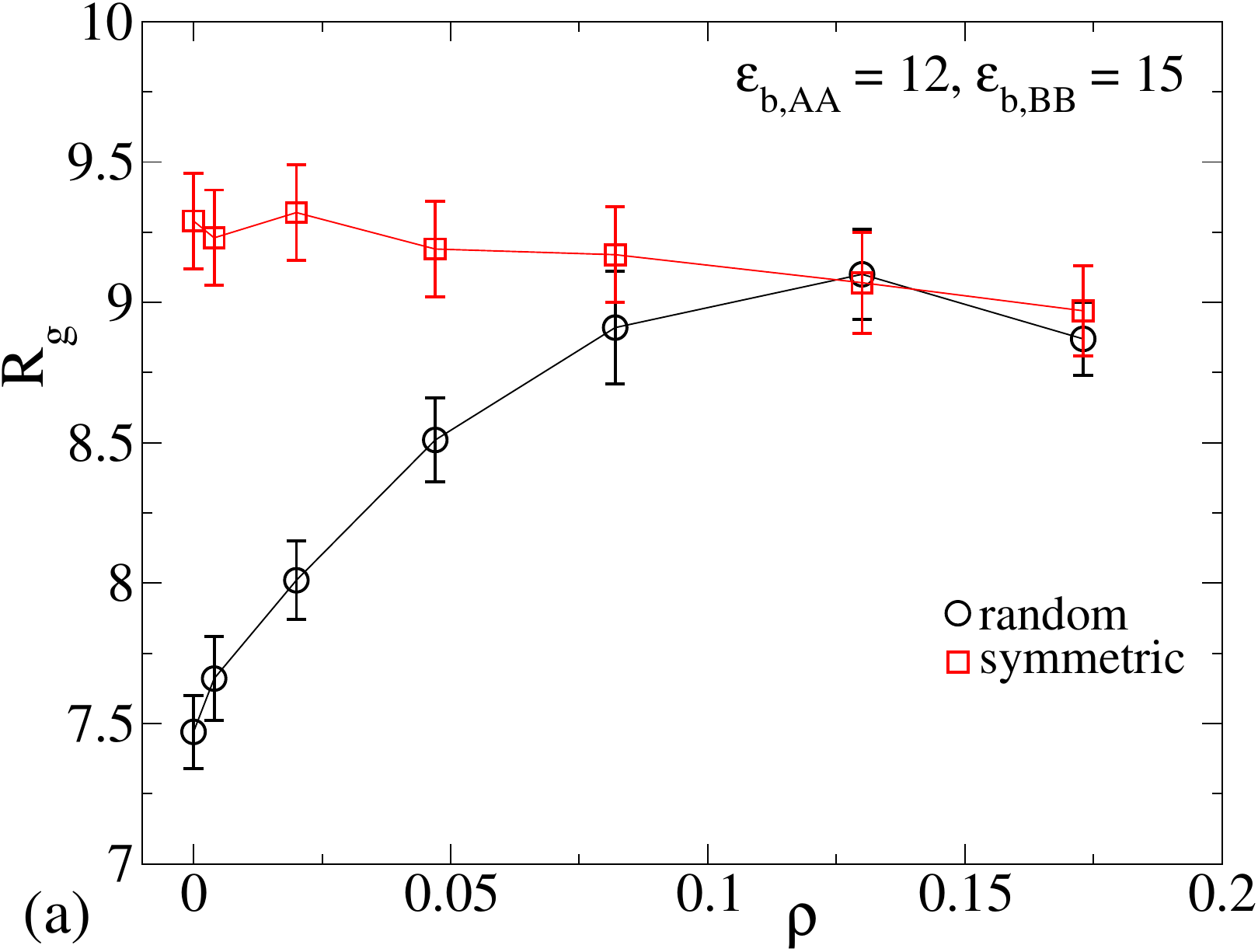}
	\\[2 mm]
   \includegraphics[width=0.45\textwidth]{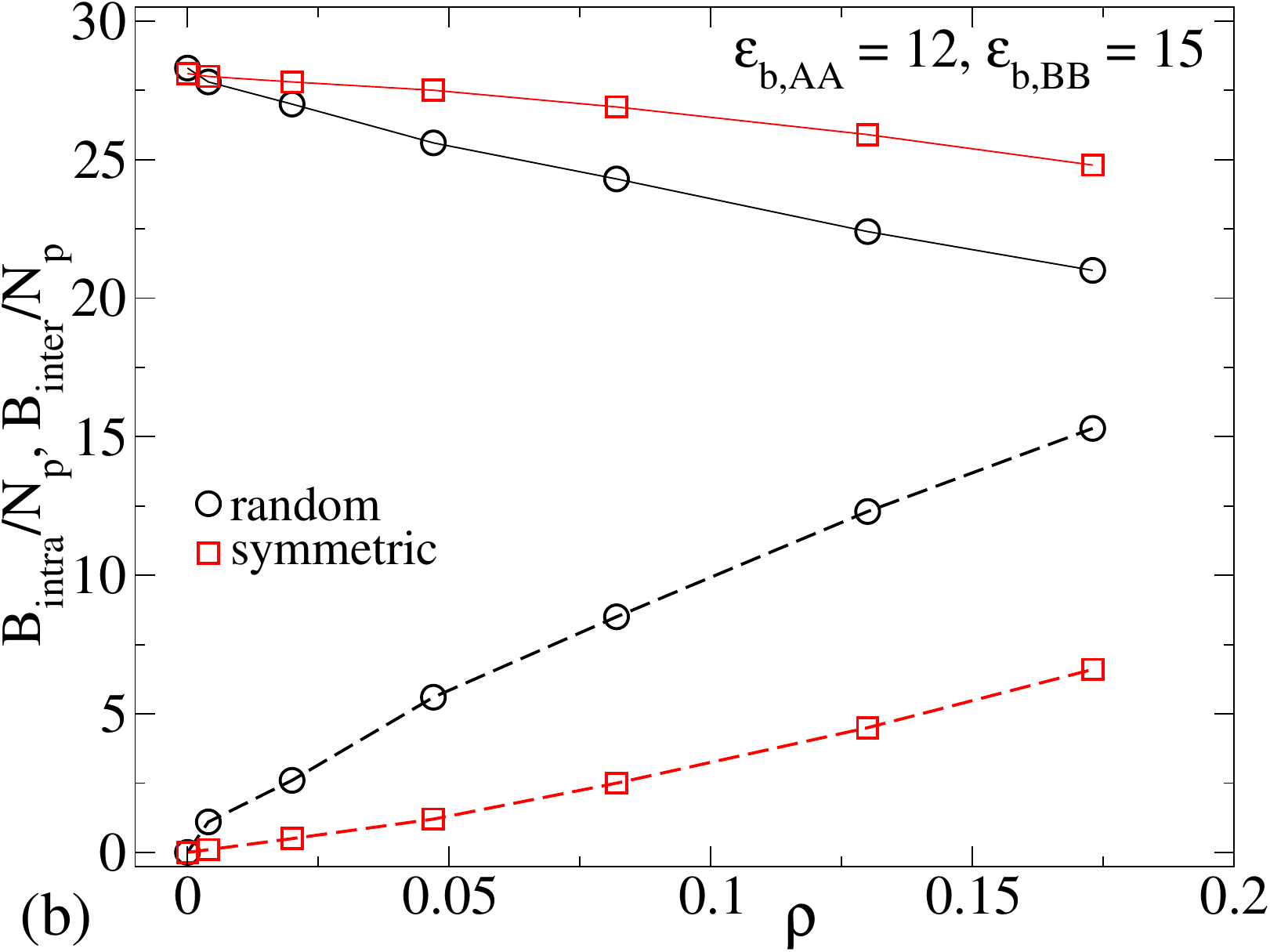}
	\\[2 mm]
    \caption{Comparison of $R_g$ and (b) intra \textit{vs.} inter-molecular bonds 
    at various densities of $x=0.5$ system for ($\epsilon_{\rm b,AA}$, $\epsilon_{\rm b,BB}$) = (12, 15). In (b)
    the intra-chain and inter-chain bonds are depicted in solid and dashed curves, respectively.}
    \label{fig:rgrandsym}
\end{figure}

\begin{figure}[h!]
	\centering
	\includegraphics[width=0.45\textwidth]{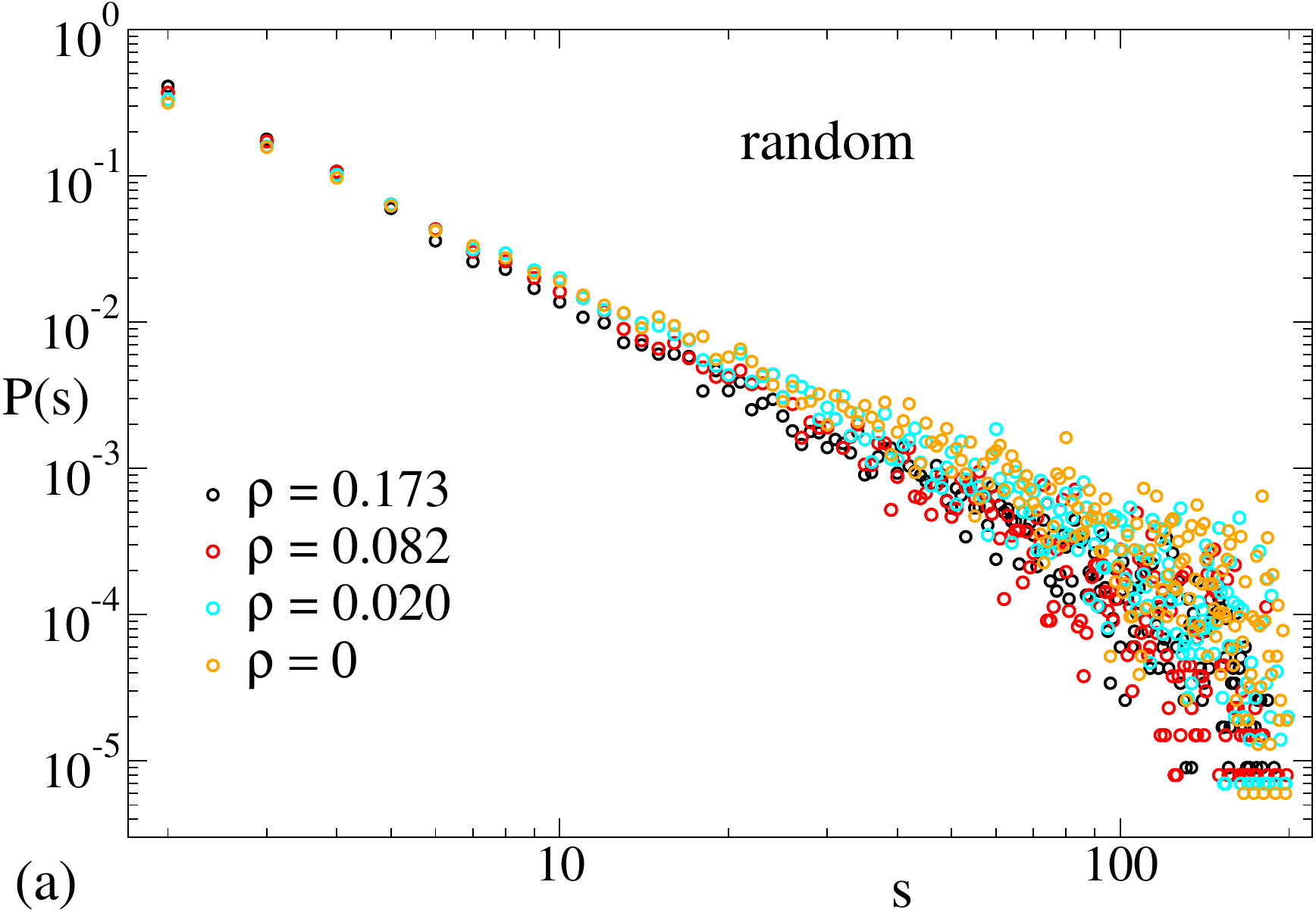}
	\\[2 mm]
	\includegraphics[width=0.45\textwidth]{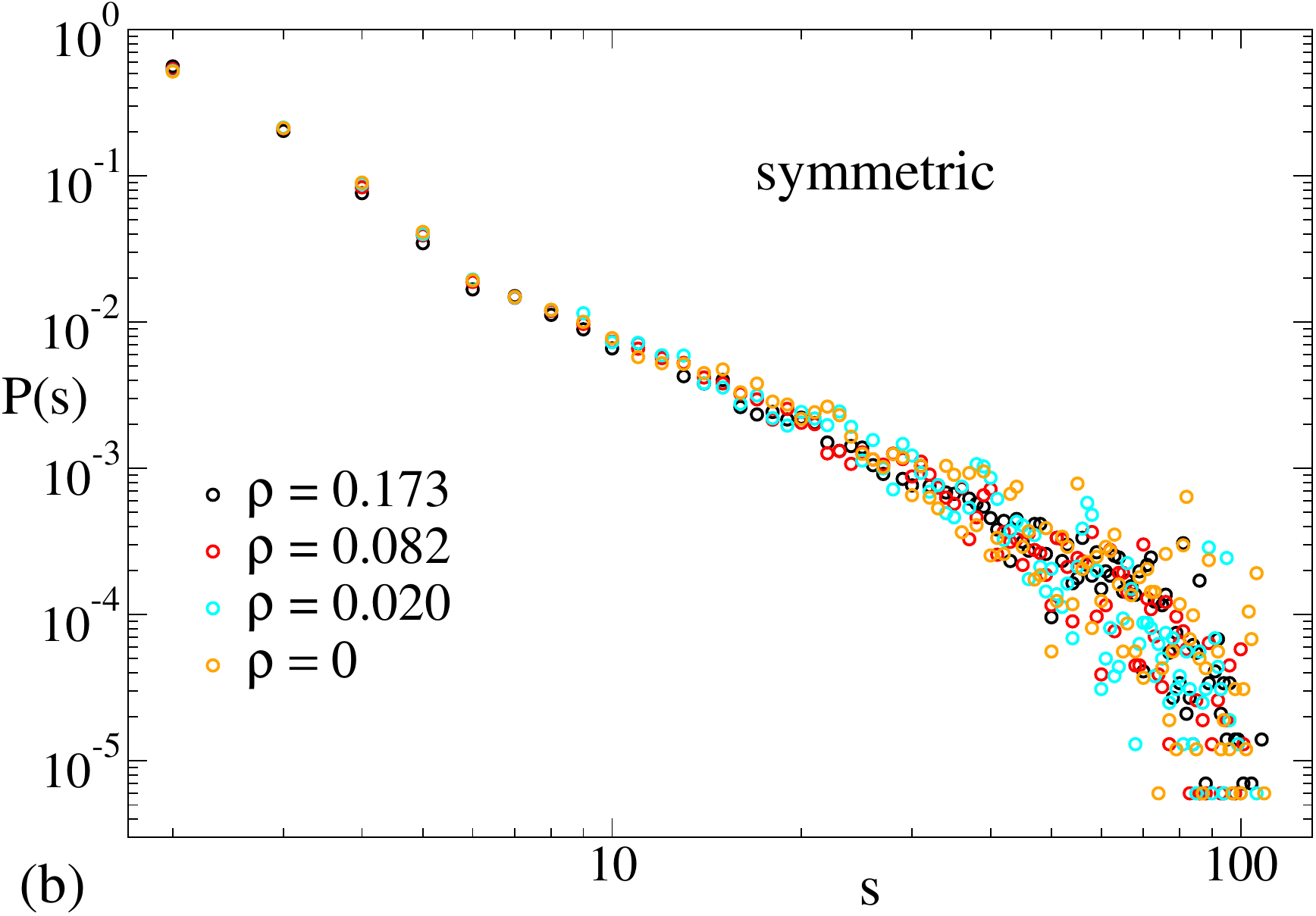}
	\\[2 mm]
	\includegraphics[width=0.45\textwidth]{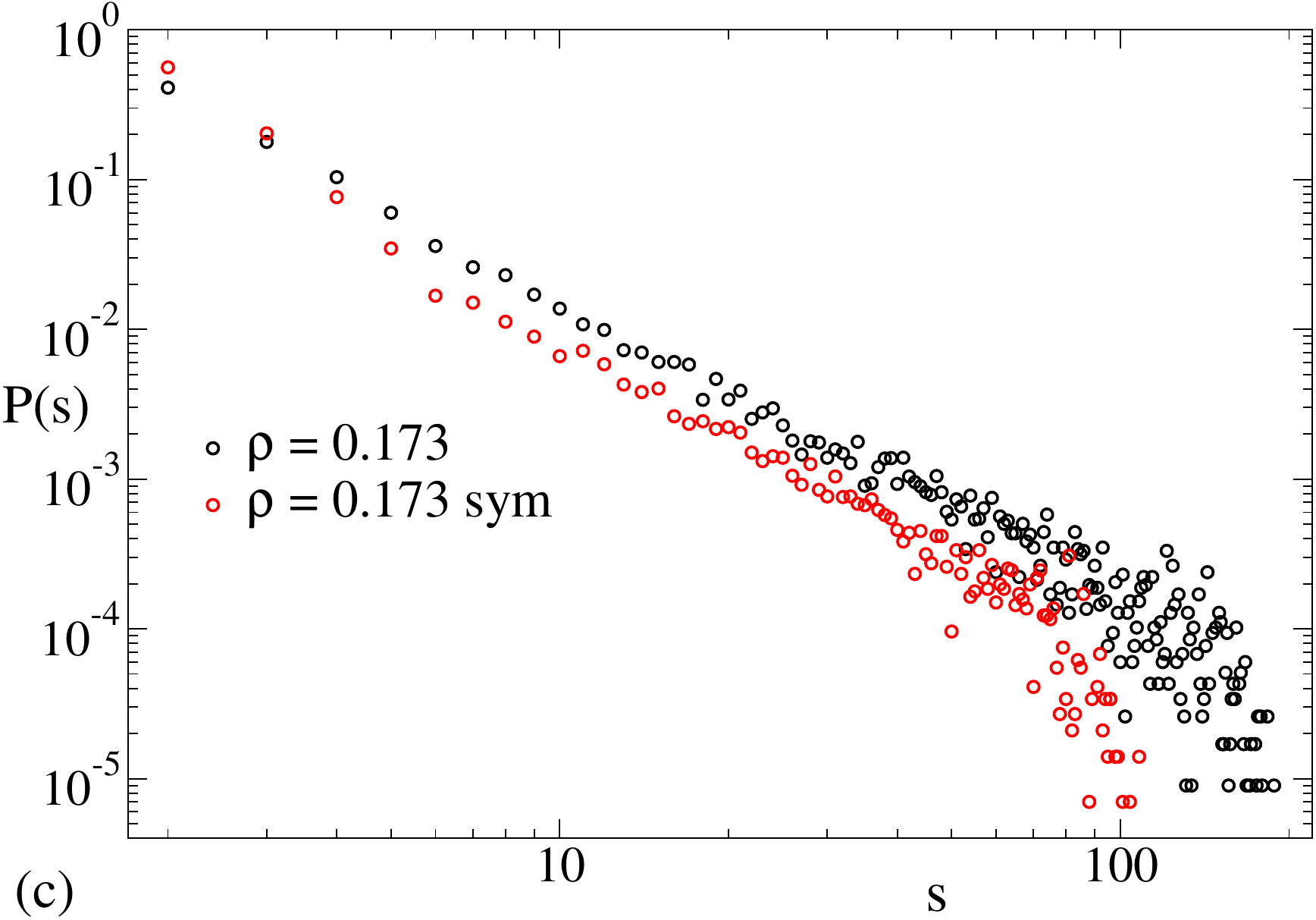}
	\\[2 mm]
	\caption{Contour distance distributions for systems studied at $x=0.5$. Here, (a), and (b) show the
	distributions for the random, and symmetric chains, respectively, as a function
        of the density. The sub figure (c) compares the profiles of the random and symmetric cases at $\rho=0.173$.}
	\label{fig:cdistrandsymm}
\end{figure}

As can be noted from Fig. \ref{fig:rgrandsym} (a), the $R_g$ of the random chains 
reduces significantly whereas that of the symmetric chains remains the same when the density is lowered. 
Further, due to the closer proximity of the reactive monomers in the symmetric case they form relatively more intra-chain bonds and less inter-chain bonds as compared to the random chains (Fig. \ref{fig:rgrandsym} (b)). 
These non-trivial trends in the size and intra \textit{vs.} inter-chain bonding affinities of the chains w.r.t 
$\rho$ can be further understood from the $P(s)$ profiles illustrated in Fig. \ref{fig:cdistrandsymm}. As we go towards 
lower densities, the probability of forming long-ranged loops increases and that of the short-ranged loops decreases 
slightly in the random case (Fig. \ref{fig:cdistrandsymm} (a)), which reflected in the decrement in their size. In 
contrast, for the symmetric case we do not find a significant variation in the $P(s)$ profiles w.r.t. $\rho$ ((Fig. \ref{fig:cdistrandsymm} (b))), in agreement with the observation from the $R_g$ analysis (Fig. \ref{fig:rgrandsym} (a)). 
Interestingly, we find that the $R_g$ of the symmetrically distributed case is relatively larger than its 
random distribution counterpart, especially at lower densities $\rho \le 0.13$ (Fig. \ref{fig:rgrandsym} (a)). 
This is intuitively expected, as the long chain loops, which are necessary for forming compact 
SCNPs \cite{blazquez2021advances}, are not possible 
with the symmetric distribution of the monomers. Indeed, as can be noted from 
Fig. \ref{fig:cdistrandsymm} (c), the probability of forming short-range loops is more in the symmetric case, 
as compared to the random case due to the closer proximity of the reactive monomers. Furthermore, due to the absence 
of the same type of monomers on the alternate block of the chain, the probability of forming the loops with $s >100$ 
is zero in the symmetric case (Fig. \ref{fig:cdistrandsymm} (c)). Therefore, due to the absence of these 
long-ranged loops and the increased number of short-ranged loops when compared with the random case, we find the 
symmetric chains to be less compact, especially so at lower densities. 

Consistently, the form factor exponent in the fractal regime for the symmetric case is higher 
($\nu \sim 0.56$) than its random counterpart ($\nu \sim 0.46$) at infinite dilution (Fig. S3 (a)), suggesting the 
extended conformations of the symmetric chains. At higher concentrations, the symmetric 
chains retain their extended conformations ($\nu \sim 0.52$, Fig. S3 (b)) due to the formation of inter-chain bonds (Fig. \ref{fig:rgrandsym} (b)).

\begin{figure}[h!]
	\centering
	\includegraphics[width=0.35\textwidth]{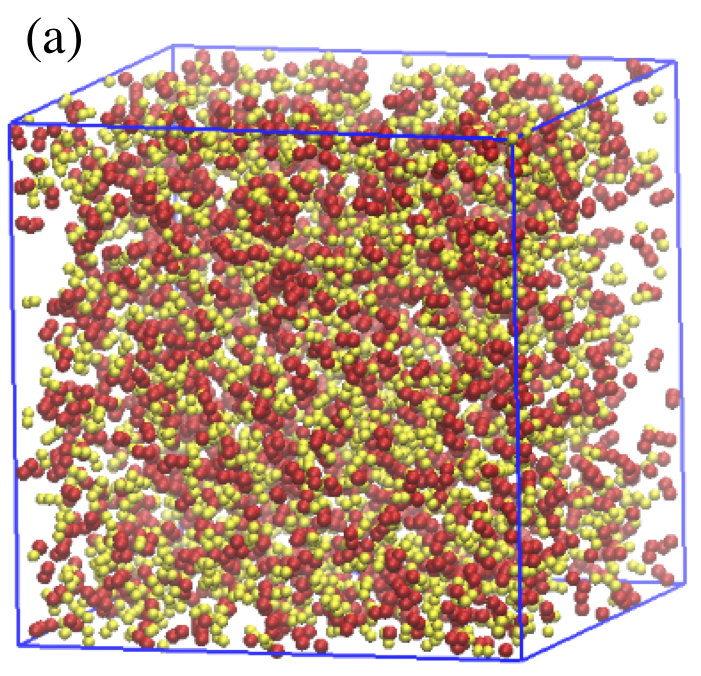}
	\\[1 mm]
	\includegraphics[width=0.35\textwidth]{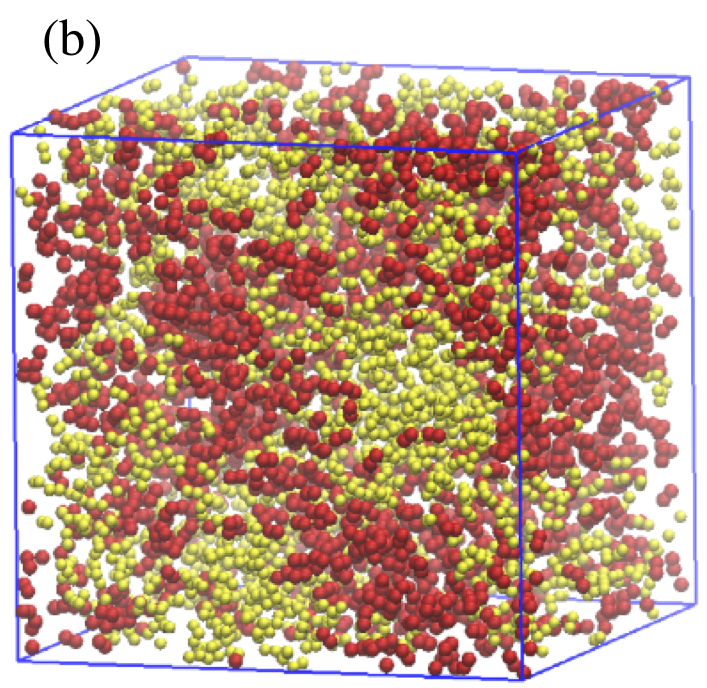}
	\\[2 mm]	
	\caption{Equilibrium snapshots of the systems at $x=0.5$ and $\rho=0.173$, illustrating the arrangement of the reactive monomers for (a) random and (b) symmetric cases. 
		A and B monomers can be distinguished in red and yellow colors, respectively.}
	\label{figrandsymm}
\end{figure}

\begin{figure}[h!]
	\centering
	\includegraphics[width=0.5\textwidth]{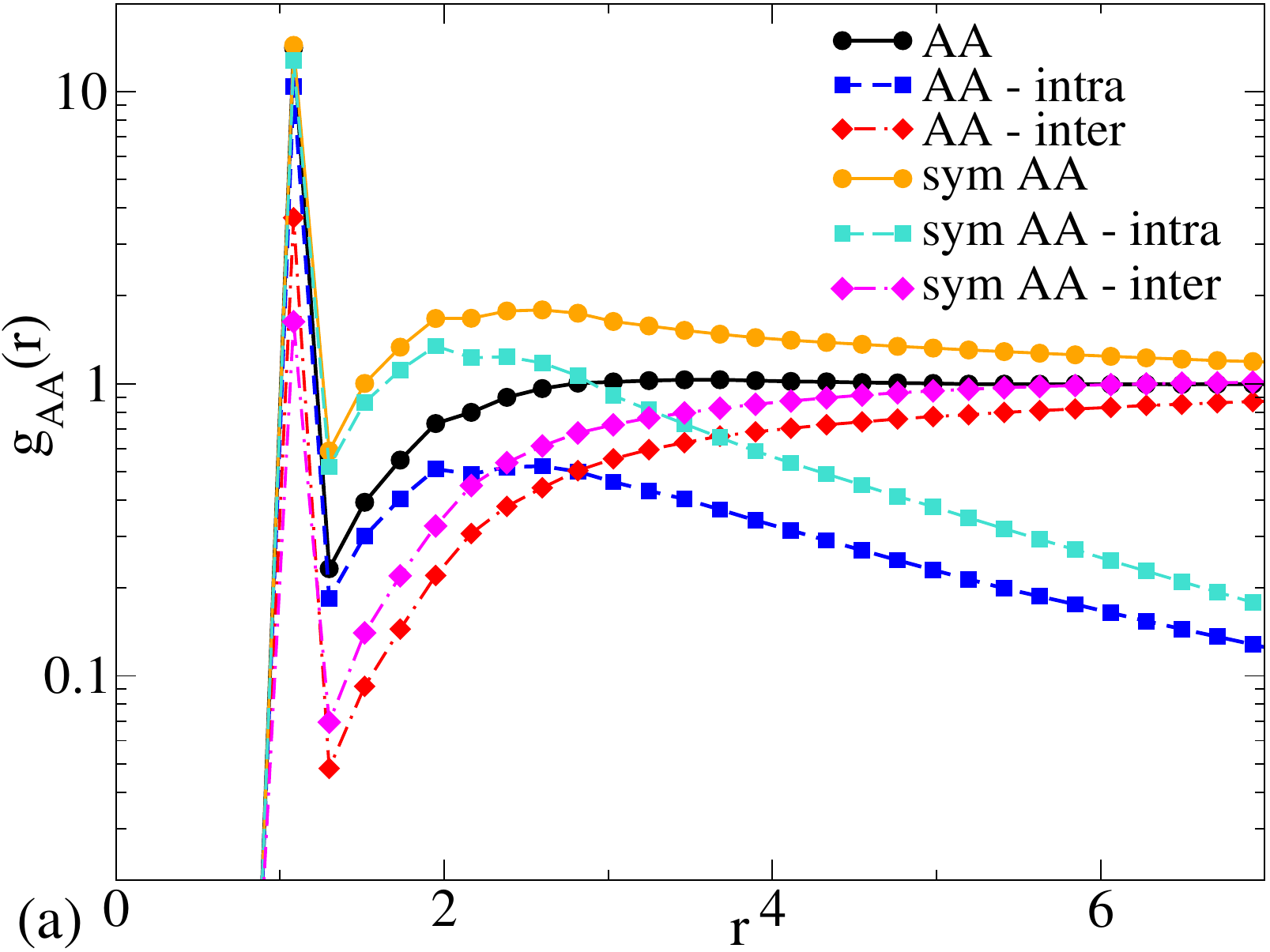}
        \\ [2 mm]
        \includegraphics[width=0.5\textwidth]{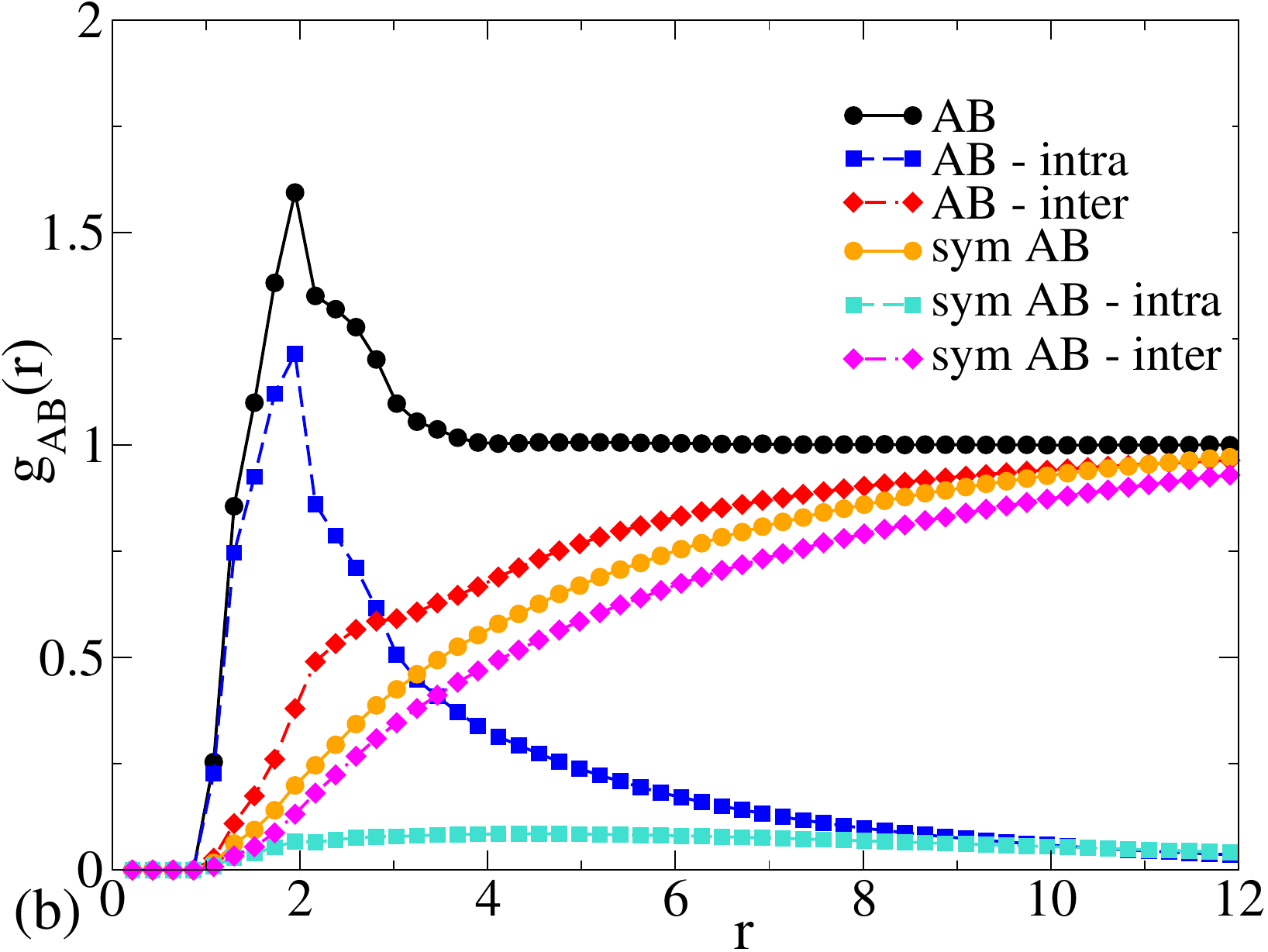}
        \\ [2 mm]
	\caption{Comparison of the partial radial distribution functions of (a) A-A, (b) A-B monomers between random and symmetric cases at $x=0.5$ and $\rho=0.173$. (a) is plotted in log-linear scale.}
	\label{figrdf}
\end{figure}

We can further understand the relative arrangement of the reactive monomers for random \textit{vs.} symmetric cases 
from the equilibrated instantaneous snapshots of the systems as illustrated in Fig. \ref{figrandsymm} at $\rho=0.173$. 
As can be seen from the instantaneous snapshots (Fig. \ref{figrandsymm}), the reactive monomers of a given type (A (red monomers) or B (yellow monomers)) have relatively stronger correlation in the symmetric case as compared to that in the random distribution. We qunatify these correlations by computing the partial radial distribution functions as illustrated in Fig. \ref{figrdf}. While the first peak position and height (at $r \sim 1$ (in $\sigma$ units)) in $g_{\rm AA}(r)$ for the symmetric chains is not different from that of the random chains,  we notice a second
peak at $r \sim 2.5 $ (in $\sigma$ units) in the symmetric case. This second peak in $g_{\rm AA}(r)$ for the symmetric chains is a signature of the increased correlations of akin reactive monomers, due to the probability of forming more short-ranged loops (Fig. \ref{fig:cdistrandsymm} (c)) and less number of inter-molecular bonds (Fig. \ref{fig:rgrandsym} (b) as argued in the above paragraphs. Although not shown here, these increased correlations in the symmetric chains are observed in the $g_{\rm BB}(r)$ profiles also. To understand the origin of the correlations better, we computed the intra-chain and inter-chain contributions 
to the g(r) function. As can be noted from Fig. \ref{figrdf} (a), the majority of the contribution to the peaks in $g_{\rm AA}(r)$ comes from the 
intra-chain correlations for both the distributions. Interestingly, we can notice that the inter-chain A-A correlations are 
stronger in the random case as compared to the symmetric case (Fig. \ref{figrdf} (a)). Further, the random case shows 
a short-range A-B correlation at $r \sim 2$, which is absent 
in the symmetric case (Fig. \ref{figrdf} (b)). This observation is a consequence of the long ranged clustering of a given type of reactive monomers (A-A and B-B) in the symmetric case. 

\subsection{Dynamical properties}
\begin{figure}[h!]
		\centering
		\includegraphics[width=0.48\textwidth]{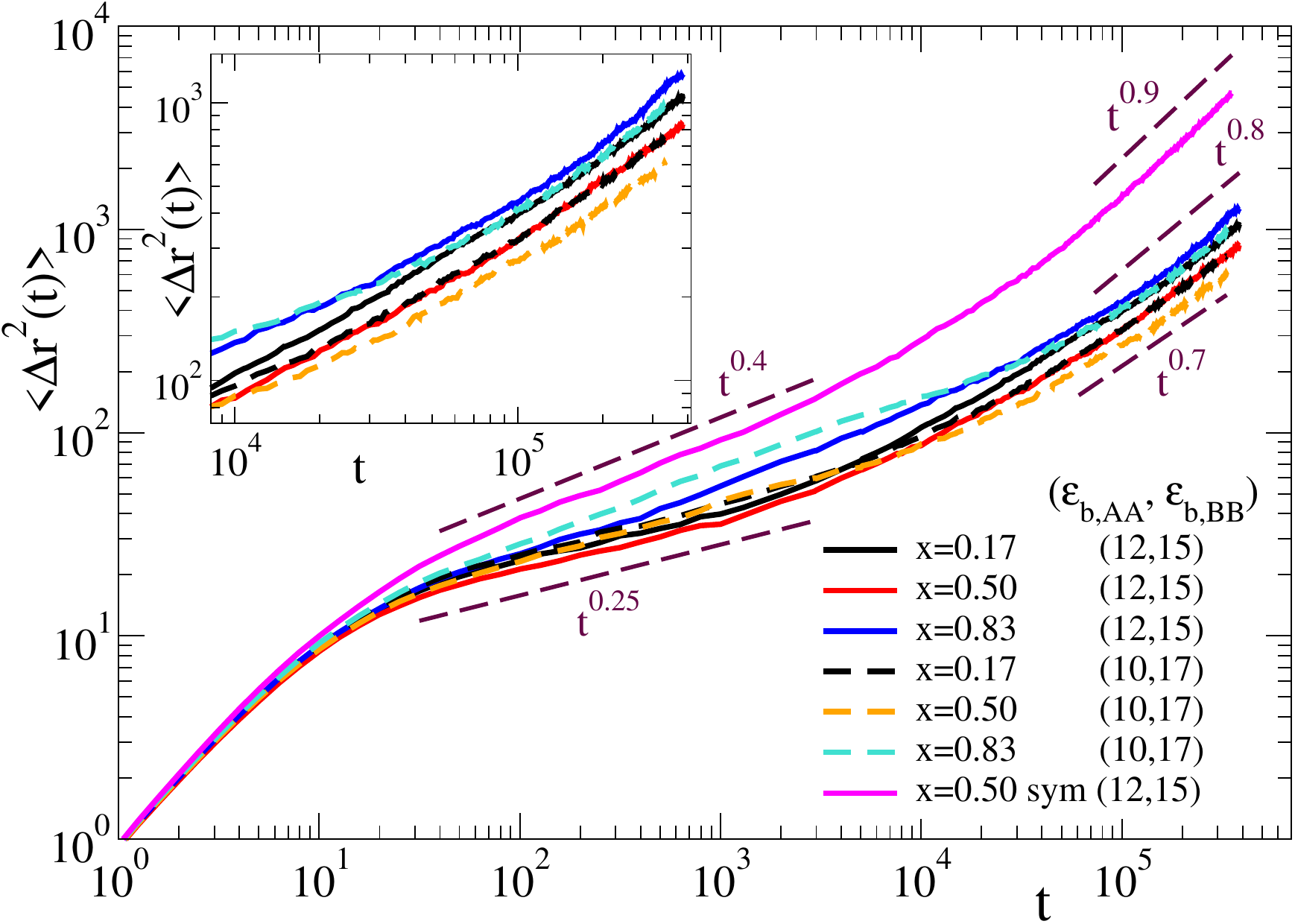}
		\\[2mm] 
		\includegraphics[width=0.48\textwidth]{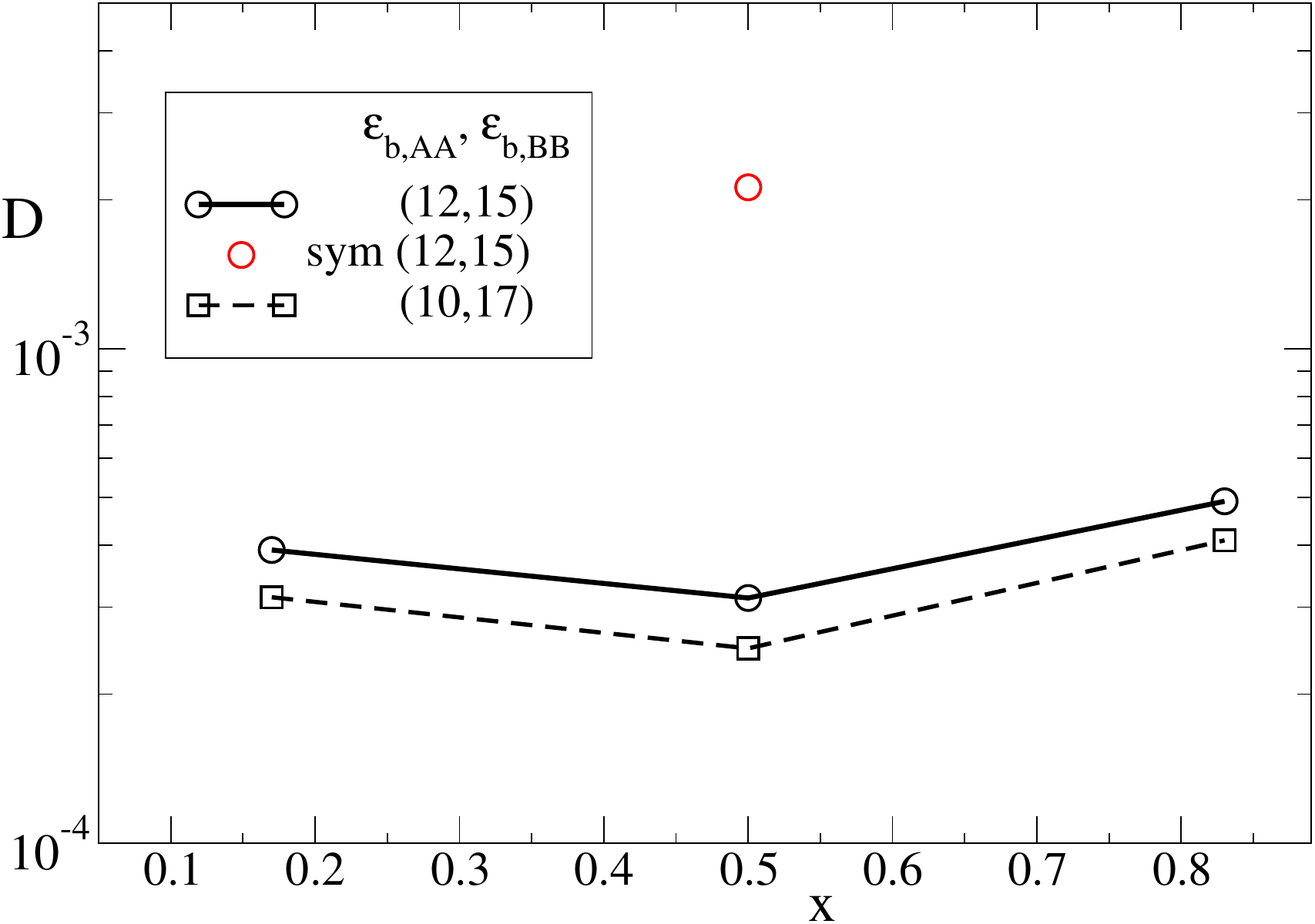}
		\\[2mm]
	\caption{(a) Monomer mean-squared displacement and (b) diffusion coefficients for the cases studied. The inset in (a) is presented for a better visibility of the MSDs corresponding to the random distribution cases studied, beyond $t=10^4$.}
	\label{figdif}
\end{figure}

Another motivation of our work is to understand how bond energy disparity, composition and the distribution of weak \textit{vs.} strong bonds affect
the dynamics and the stress relaxation of the network. To this end, we first discuss the mean-squared displacement (MSD) of the 
individual monomers defined as, $\text{MSD}(t)=\langle {(\vec{r}_{i}(t) - \vec{r}_{i}(0))}^2\rangle$.  As illustrated in 
Fig. \ref{figdif} (a), at shorter timescales the dynamics is equivalent across all the systems. However, at the intermediate 
timescales, the plateau regime can be observed, where the localization length, $\Delta = \sqrt{MSD} \approx 5-9 $ is of the order
of the size of the polymer. Such a large localization length and the slowed
down dynamics in this plateau regime (MSD $\sim t^\delta$ with $\delta$ ranging from 0.25 to 0.4 across different systems 
considered) is a signature of the system spanning network formed by these reversible SCNPs \cite{formanek2021gel}. 

Interestingly, at any given ($\epsilon_{\rm b,AA}, \epsilon_{\rm b,BB}$), the dynamics is 
non-monotonic in nature w.r.t $x$, as we can note that the trend in the diffusion coefficient ($D$), obtained from the long timescale 
linear fitting of the MSD (Fig. \ref{figdif} (b)) follows $D(x=0.83) > D(x=0.17) > D(x=0.5)$. Surprisingly, 
although the bonding probabilities of the symmetric ($p_B=0.94$) and the random chains ($p_B=0.95$) are similar at $x=0.5$, 
the symmetric chains are the relatively more diffuse (Fig. \ref{figdif}). We can further understand these trends in the chain dynamics as 
a function of $x$ and monomer distribution, from the $B_{\rm inter}/N_p$ (Figs. S1 (c) and 4 (b)). Clearly,
the number of inter-molecular bonds are higher at $x=0.5$ as compared to $x=0.17$ and $0.83$ (random sequences, Fig. S1 (c)
and the discussion in section 3.1.2). Therefore, one can expect the dynamics to be slower in $x=0.5$ case as illustrated in
Fig. \ref{figdif} (b). Also, as discussed in section 3.2, the number of inter-chains bonds are significantly lower for 
the symmetric sequences when compared with that of the random sequences at $x=0.5$ (Fig. \ref{fig:rgrandsym} (b)). Moreover,
the connectivity of the symmetrically distributed chains is almost half of that of the randomly distributed chains (Fig. S4), making 
the dynamics faster in the symmetric case. A similar observation was reported by Ricarte et al \cite{ricarte2021unentangled} 
where-in the zero-shear viscosity of a model vitrimer system was found to be more for a random sequence than for a 
blocky sequence.

It is noteworthy to mention that although the number of inter-chain bonds (Fig. S1 (c)) and the connectivity per chain 
(Fig. S4) are comparable in $x=0.17$ and $x=0.83$ cases, we find the dynamics to be faster in $x=0.83$ case. Further, at 
the intermediate timescales, the dynamics is faster in ($\epsilon_{\rm b,AA}, \epsilon_{\rm b,BB}$) = (10-17) case as compared 
to the (12-15) case. However, at larger timescales, the dynamics is faster in 
($\epsilon_{\rm b,AA}, \epsilon_{\rm b,BB}$) = (12-15) system at any given reactive monomer composition. To understand the 
origin of these crossovers in the dynamics of the two bond energy disparity cases studied, we computed the bond auto-correlation functions as described below. 

\begin{figure}[h!]
	\centering
	\includegraphics[width=0.47\textwidth]{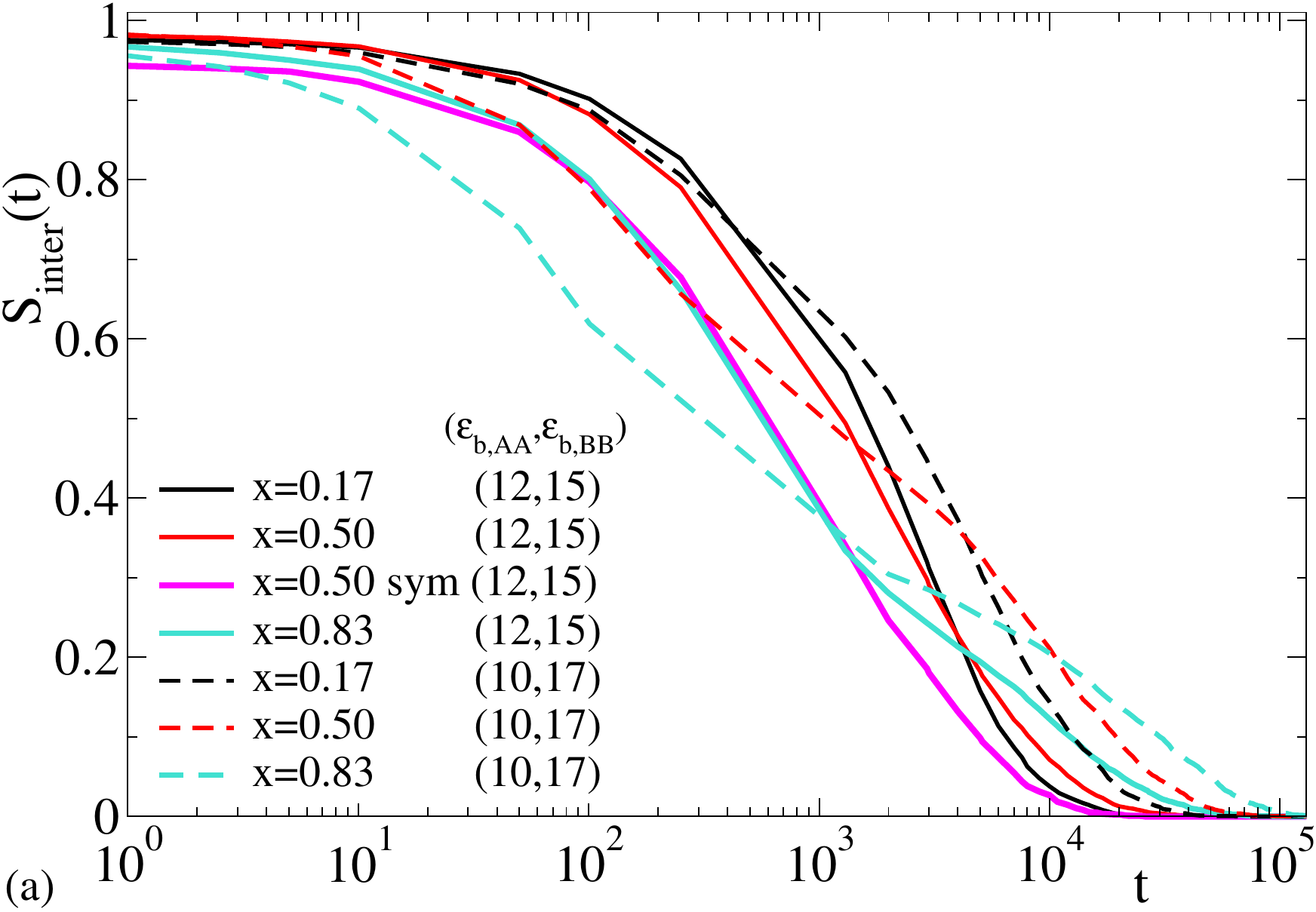}
	\\[2mm]
	\includegraphics[width=0.47\textwidth]{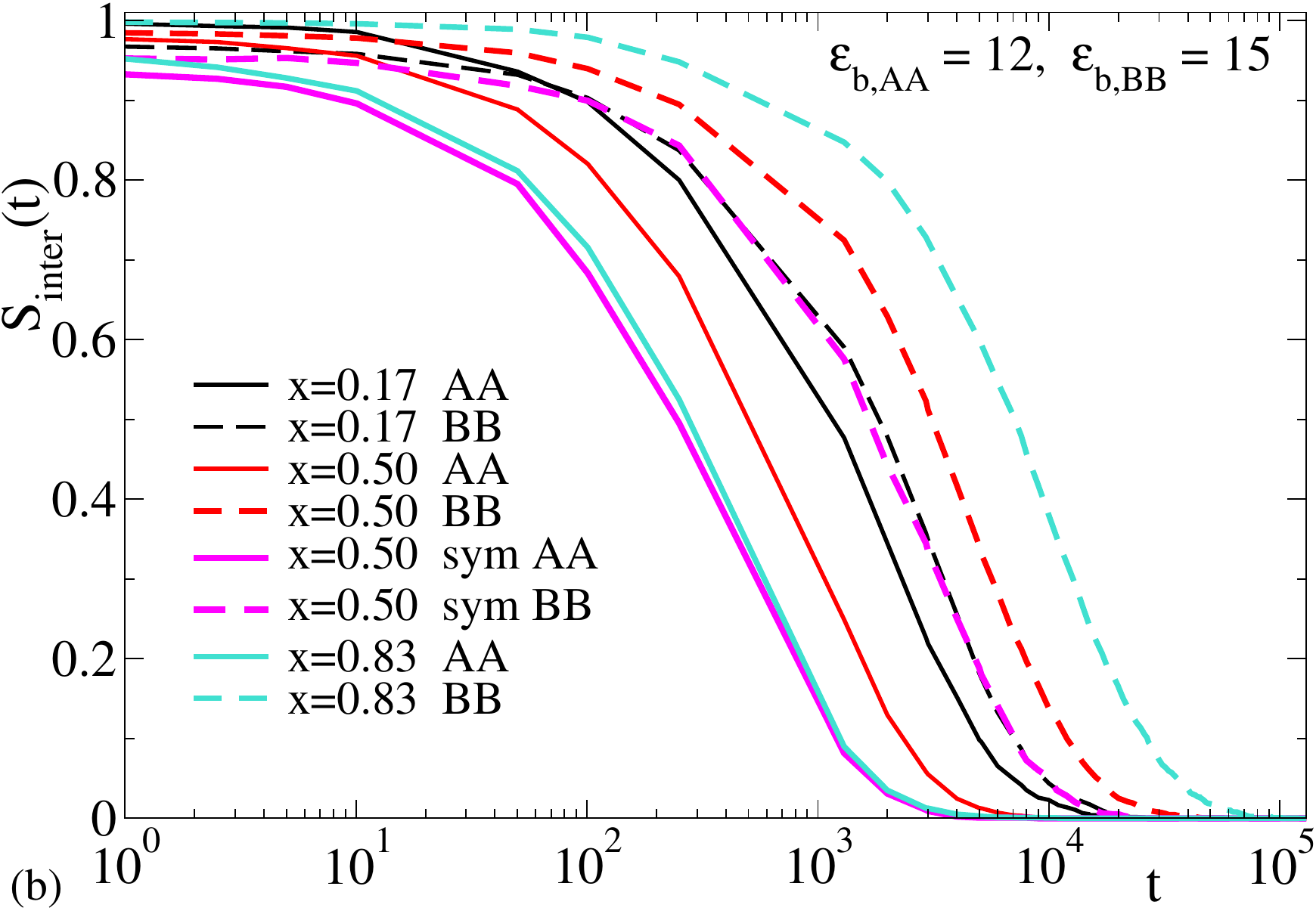}
	\\[2mm]
	\includegraphics[width=0.47\textwidth]{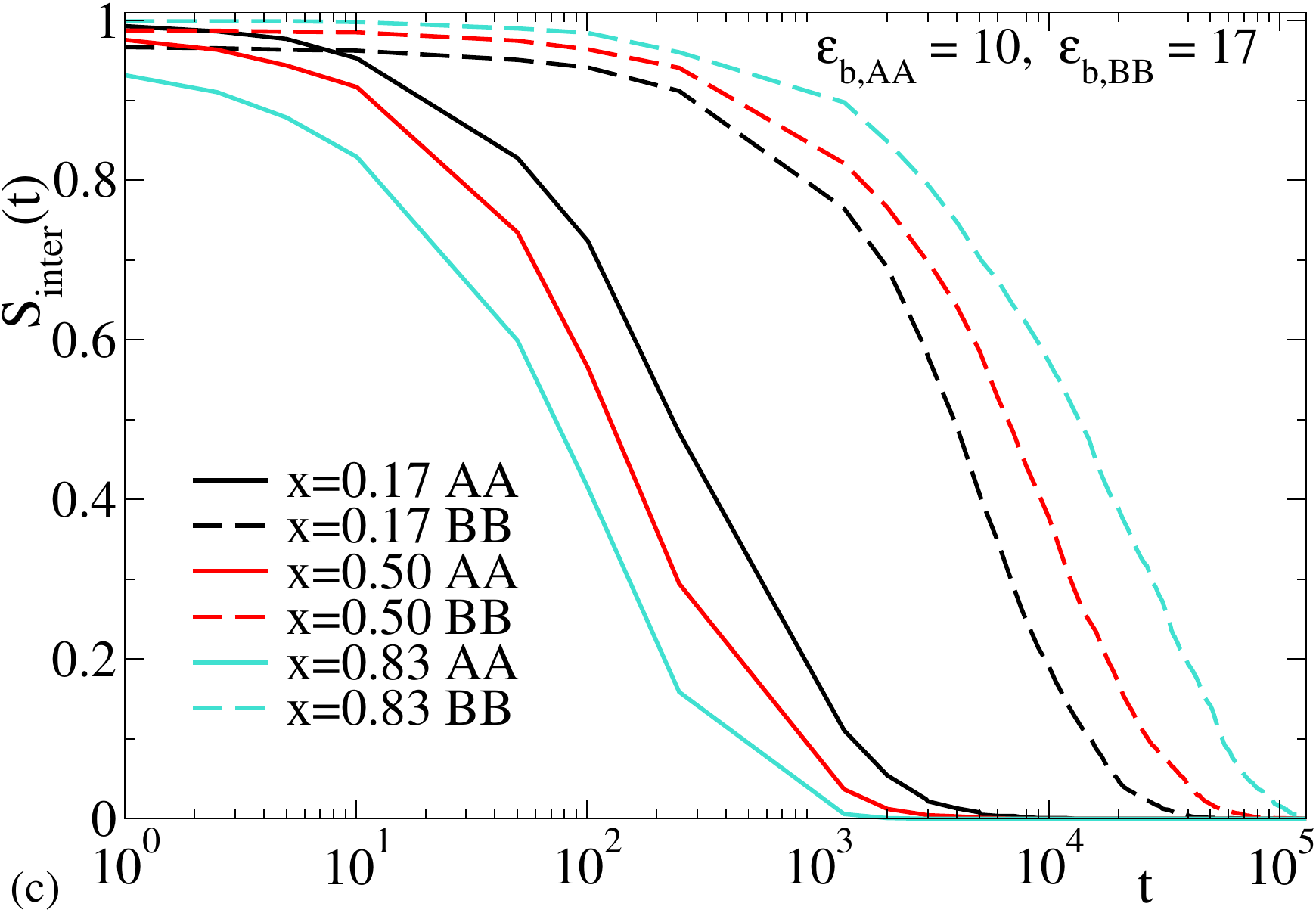}
	\\[2mm]	
	\caption{(a) Total inter-molecular bond correlations of the chains at $\rho=0.173$. Contributions to the total inter-bond correlation functions
	from the A-A and B-B bonds for (12-15) and (10-17) cases are shown in (b), and (c), respectively.}
	\label{bondcorr}
\end{figure}

In a system where a system spanning network is persistent, the dynamics is dependent on the
formation and breaking of inter-molecular bonds \cite{formanek2021gel}. Therefore, we study the dynamics of the inter-chain bonds 
by calculating the bond self-correlation function, which quantifies the life-time of the inter-molecular bonds between the 
reactive monomers. We define the self-correlation function as follows,
\begin{equation}
\label{sinter}
S_{\text{inter}}(t) = \sum_{i,j=reac.\,\,mon.}\frac{\langle B_{ij}(t) B_{ij}(0)\rangle}{{\langle B_{ij}(0) \rangle}^2},
\end{equation}
where $B_{ij}(t)=1$ if there is a bond between the reactive monomers $i$ and $j$ for all the times from 0 to $t$. If
the bond is broken at any time earlier to $t$, then $B_{ij}(t)=0$ for that reactive monomer pair $i$, $j$. By 
construction, the indices $i$ and $j$ run over all the reactive monomer pairs. Clearly, the total (A + B) inter-bond lifetimes
($\tau$, defined as the time where the $S_{\rm inter}(t)$ falls to $1/e$) are larger in the $x=0.17$ case over the $x=0.83$ at
both the bond energy disparities studied (Fig. \ref{bondcorr} (a)). As a consequence, the dynamics in the $x=0.83$ case is faster 
as compared to $x=0.17$. Therefore, the number of inter-chain bonds and their lifetimes are two crucial aspects in understanding the 
dynamical behavior of the networks formed. In addition, as shown in Fig. \ref{bondcorr} (a), at intermediate times the bonds are 
better correlated in $(\epsilon_{\rm b,AA}, \epsilon_{\rm b, BB})$ = (12-15) case than those in (10-17) case for all the $x$ 
values simulated. However, at longer-times the $(\epsilon_{\rm b,AA}, \epsilon_{\rm b, BB})$ = (10-17) case dominates in the 
bond correlation, making the overall bond survival times larger in this case. 

To understand the role of weak \textit{vs.} strong bonds in tuning the bond correlations at different timescales, we further 
calculated the contributions from A-A and B-B bonds to $S_{inter}$ for $(\epsilon_{\rm b,AA}, \epsilon_{\rm b, BB})$ = (12-15) 
and (10-17) cases, as shown in Fig. \ref{bondcorr} (b) and (c), respectively. Firstly, the correlation times of the stronger bonds 
($\tau_{\rm B}$) are greater than that of weaker bonds ($\tau_{\rm A}$), as expected. Larger the bond energy disparity 
between the weak and strong bonds, the greater is the difference between $\tau_{\rm B}$ and $\tau_{\rm A}$. One would expect the 
bond correlation times of a given species (A or B) to be monotonically increasing as we increase their relative fraction. 
However, we find the trend in the correlation times to be decreasing w.r.t. the fraction of a given species. For instance, the correlation times for the weaker 
bonds follow $\tau_{\rm A}(x=0.17) > \tau_{\rm A}(x=0.5) > \tau_{\rm A}(x=0.83)$, even though the trend in the number of 
A-A inter-chain bonds is $B_{\rm inter, A}(x=0.17) < B_{\rm inter, A}(x=0.5) < B_{\rm inter,A}(x=0.83)$ (see Fig. S1 (c)). 
This implies that the more the number of species forming the bonds (weak or strong), more is the number of bond exchanges 
occurring in the system. The same trends apply to the B species as well. Moreover, the weaker A-A bonds die out faster 
in $(\epsilon_{\rm b,AA}, \epsilon_{\rm b, BB})$ = (10-17) case as compared to (12,15) case. (Fig. \ref{bondcorr} (b) and (c)) However, at longer times because of the existence of the relatively stronger B-B bonds in 
$(\epsilon_{\rm b,AA}, \epsilon_{\rm b, BB})$ = (10-17) case, they have greater $\tau$ values, compared to that of (12-15) case.
Therefore, we can conclude that the dynamics in the intermediate times can be tuned by the strength of the weaker bonds, thus (10,17) 
system has faster dynamics as compared to (12,15) case at these timescales. Similarly, the dynamics at the longer timescales 
is determined by the strength of the stronger bonds, making the $(\epsilon_{\rm b,AA}, \epsilon_{\rm b, BB})$ = (12,15) system achieve faster terminal relaxation. 

\subsubsection{Stress relaxation}
The strength of the reversible bonds is known to tune the mechanical properties of the polymer gels from solid to liquid-like 
depending on the timescales \cite{raffaelli2021stress}. Therefore, to characterize the viscoelastic properties of our dual 
cross-linked system, we compute the stress auto-correlation function ($G(t)$) from the equilibrium MD simulations using \cite{ramirez2010efficient}
\begin{align}
G(t) &= \frac{V}{5k_B T} \left[ \langle \sigma_{xy}(t) \sigma_{xy}(0)\rangle + \langle \sigma_{yz}(t) \sigma_{yz}(0)\rangle + \langle \sigma_{zx}(t) \sigma_{zx}(0)\rangle \right] \nonumber \\
& + \frac{V}{30k_B T} \left[ \langle N_{xy}(t) N_{xy}(0)\rangle + \langle N_{yz}(t) N_{yz}(0)\rangle + \langle N_{xz}(t) N_{xz}(0)\rangle \right], \label{saf}
\end{align}
where $N_{\alpha\beta}=\sigma_{\alpha \alpha} - \sigma_{\beta \beta}$. The stress tensor $\sigma_{\alpha\beta}$ is calculated 
on the fly during the MD simulation using,
\begin{align}
\sigma_{\alpha\beta} & = -P_{\alpha\beta} \nonumber \\
& = -\frac{1}{V} \sum_{i=1}^{N} m_i v_{i,\alpha} v_{i,\beta} - \frac{1}{V} \sum_{i=1}^{N-1} \sum_{j=i+1}^{N} F_{ij,\alpha} r_{ij,\beta}, \label{stressten}
\end{align}
where $\alpha$ and $\beta$ take the values $x,y,z$. Further, we perform a running average of the raw $G(t)$ data for smoothing 
the noise, with a length of 5 data points. \cite{raffaelli2021stress}

\begin{figure}[h!]
	\centering
	\includegraphics[width=0.5\textwidth]{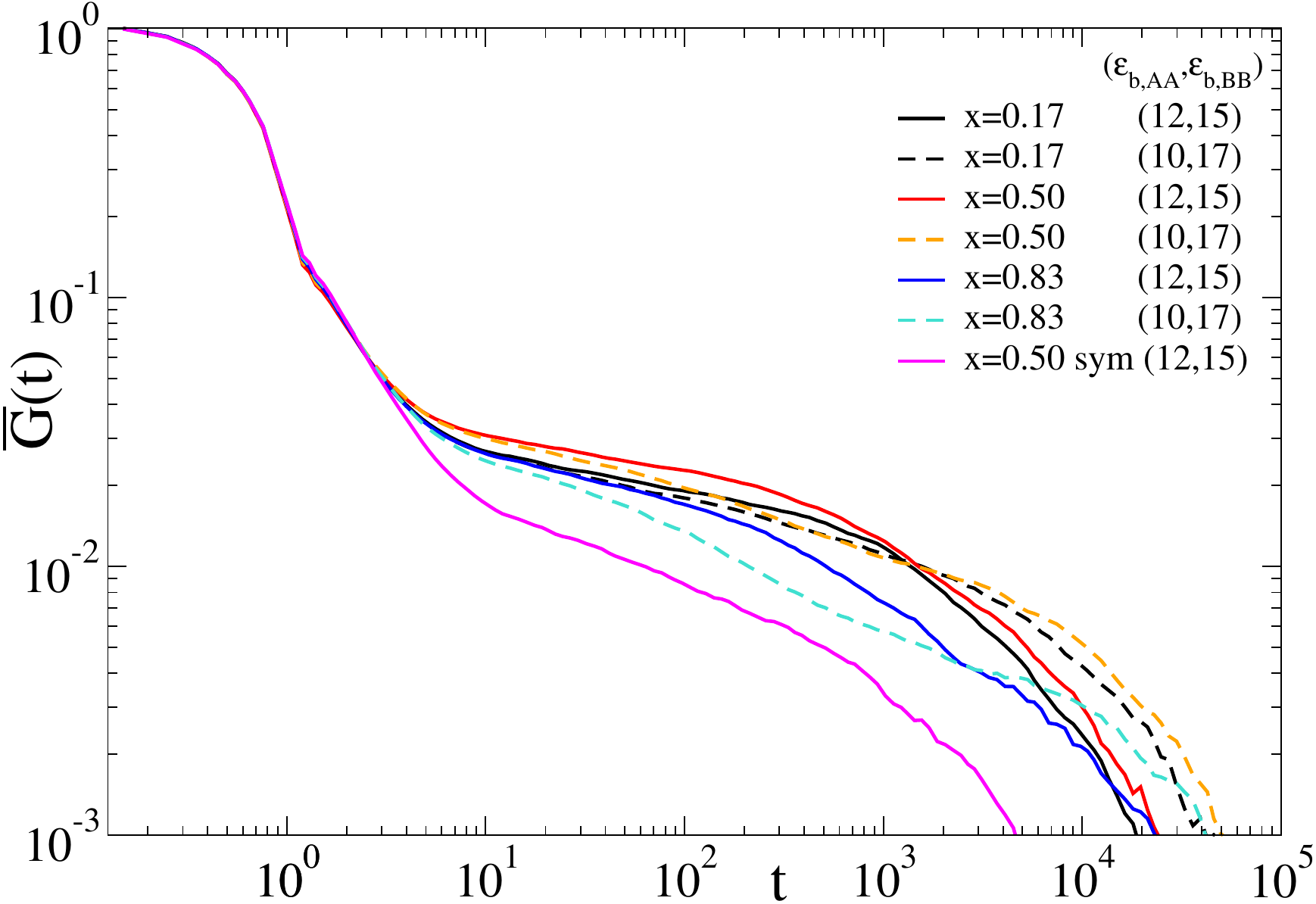}
	\caption{Stress auto-correlation function profiles for all the cases studied (running average performed over 5 data points.}
	\label{figsaf}
\end{figure}

Fig. \ref{figsaf} illustrates the time evolution of stress relaxation modulus ($\overline{G}(t)$) 
averaged over the 8 independent replicas for all the systems studied (Table: \ref{tab:1}). Firstly, all the observed 
$\overline{G}(t)$ profiles suggest a typical response of an RPN, wherein there exists a plateau regime followed by a 
terminal relaxation. \cite{ciarella2018dynamics} At short timescales ($t \le 5$) the relaxation behavior is indistinguishable
across all the cases studied. At the intermediate timescales (from $t \sim 5$ till about $t \sim 10^3$), we find a plateau regime,
which signifies the onset of the network formation. The weaker bonds relax at these intermediate timescales, as argued in the $S_{inter}(t)$ discussion. Beyond $t \sim 10^3$, the stronger bonds also start relaxing and thus we observe the terminal 
relaxation of the stress modulus at these long timescales. 
As expected, the terminal relaxation times are inversely proportional to the 
diffusion constant $D$ (Fig. \ref{figdif} (b)), \textit{i.e.,} the stress relaxes faster in the system that is more diffusive. 
Consistent with the trends obtained from the MSD analysis (Fig. \ref{figdif}) and the 
total bond correlation times (Fig. \ref{bondcorr}), the stress modulus is greater in 
$(\epsilon_{\rm b, AA}, \epsilon_{\rm b, BB})$ = (12-15) case as compared the that in (10-17) case, 
due to the relatively slower relaxation of the weaker bonds, at the intermediate time scales ($t \le 10^3$) for a given 
value of $x$. Similarly, at longer time scales ($t > 10^3$), the relaxation behavior is determined by the strength of 
the stronger bonds, thus the stress relaxes faster in $(\epsilon_{\rm b, AA}, \epsilon_{\rm b, BB})$ = (12-15) system. 
At $x=0.5,$ the network with the symmetric distribution of the chains exhibits the fastest stress relaxation and a 
poor plateau modulus, as compared to the system of random distribution of the reactive monomers, due to the reduction in the 
number of inter-chain bonds, as highlighted above. Further, the non-monotonicity in the dynamics of the monomers w.r.t. $x$,
is also observed here in the $\overline{G}(t)$ at both the intermediate and long timescale regimes. Therefore, we can tune
the viscoelastic response of the networks through the strength, composition, and the distribution of the strong and weak bonds in a dual crosslinked RPN.

\section{Conclusions}
We presented a thorough investigation of the structural and dynamical aspects of the reversible SCNPs and their networks 
made of dual crosslinkers, using equilibrium Langevin dynamics simulations. Specifically, we probed the influence of the 
composition, bond energy disparity, and the distribution of the weak \textit{vs.} strong bonds on the size, intra-molecular 
\textit{vs.} inter-molecular bonding probabilities, and the molecular conformations of the SCNPs. At infinite dilution, 
we find that the dual crosslinked SCNPs to be compact as compared to the singly crosslinked ones, confirming the 
effectiveness of orthogonal crosslinking in obtaining globular morphologies of the SCNPs. Further, at infinite dilution,
we find that increasing the contour distance between alike reactive monomers (A or B) results in the formation of more 
long-ranged loops. Though this led to a slight non-monotonicity in the size of the SCNPs w.r.t $x$, we do not observe any 
significant change in the internal structure of the SCNPs. Similarly, at the finite concentration ($\rho=0.173$), we find
the sizes of the chains to be significantly larger as compared to $\rho=0$ case, due to the formation of extended 
conformations through inter-molecular bonds. 

For the cases considered, we find that the total bonding probability of the SCNPs is mainly dependent on the fraction and 
strength of the weaker bonds, at both the concentrations studied. Thus, we find that the higher the bond energy disparity 
between the orthogonal monomers, the lesser the probability of forming weaker bonds. As a consequence, the system with 
lesser bond energy disparity ($(\epsilon_{\rm b,AA}, \epsilon_{\rm b,BB}) = (12,15)$) has a higher number of bonds (intra- 
and inter-molecular) formed at both the concentrations studied. Further, we find that the symmetric arrangement of the 
reactive monomers gives rise to an extended conformation of the SCNP, as compared to its random counterpart owing to the 
formation of more short-ranged loops, at lower densities. Although, the conformations of the symmetric and 
random chains are extendend at higher concentrations, the symmetric chains participate in comparatively lesser (greater) 
number of inter-chain (intra-chain) bonds. As a consequence, the correlation between the reactive monomers of a given type 
is more in the symmetric arrangement at higher concentration, suggesting faster bond exchanges and stress-relaxation, 
compared to the random case. 

The interplay of weak \textit{vs.} strong bonds is also evident in the dynamical properties of the network. 
At a given $x$, we find that the dynamics in the plateau regime is dependent on the strength of the weaker bonds, whereas
that in the terminal regime is controlled by the stronger bonds. Finally, we find a non-monotonic dependence in the 
dynamics of the system w.r.t. $x$, with $x=0.5$ being the slowest network, due to the formation of relatively more number
of inter-chain bonds. Thus, our study highlights the importance of the relative bond energy strengths of the crosslinking monomers in the DDPNs. Further, it highlights how a variation in the reactive monomer composition, and the distribution can tune the viscoelastic response of the networks. 

\bibliography{refs}

\newpage
\renewcommand{\thefigure}{S\arabic{figure}}
\setcounter{figure}{0} 
\noindent\textbf{\Large{Supporting Information}}
\newline
\noindent\textbf{\large{1. Intra \textit{vs.} inter-molecular bonds per chain}} 
\newline
The competition between the formation of intra and inter-molecular bonds affects  
molecular conformations and dynamics. Therefore, In Fig. \ref{nbonds} we present the number of 
intra-molecular ($B_{\rm intra}/N_p$) and inter-molecular ($B_{\rm inter}/N_p$) bonds per chain 
for the random monomer distributions studied. 

\begin{figure}[h!]
	\centering
		\includegraphics[width=0.45\textwidth]{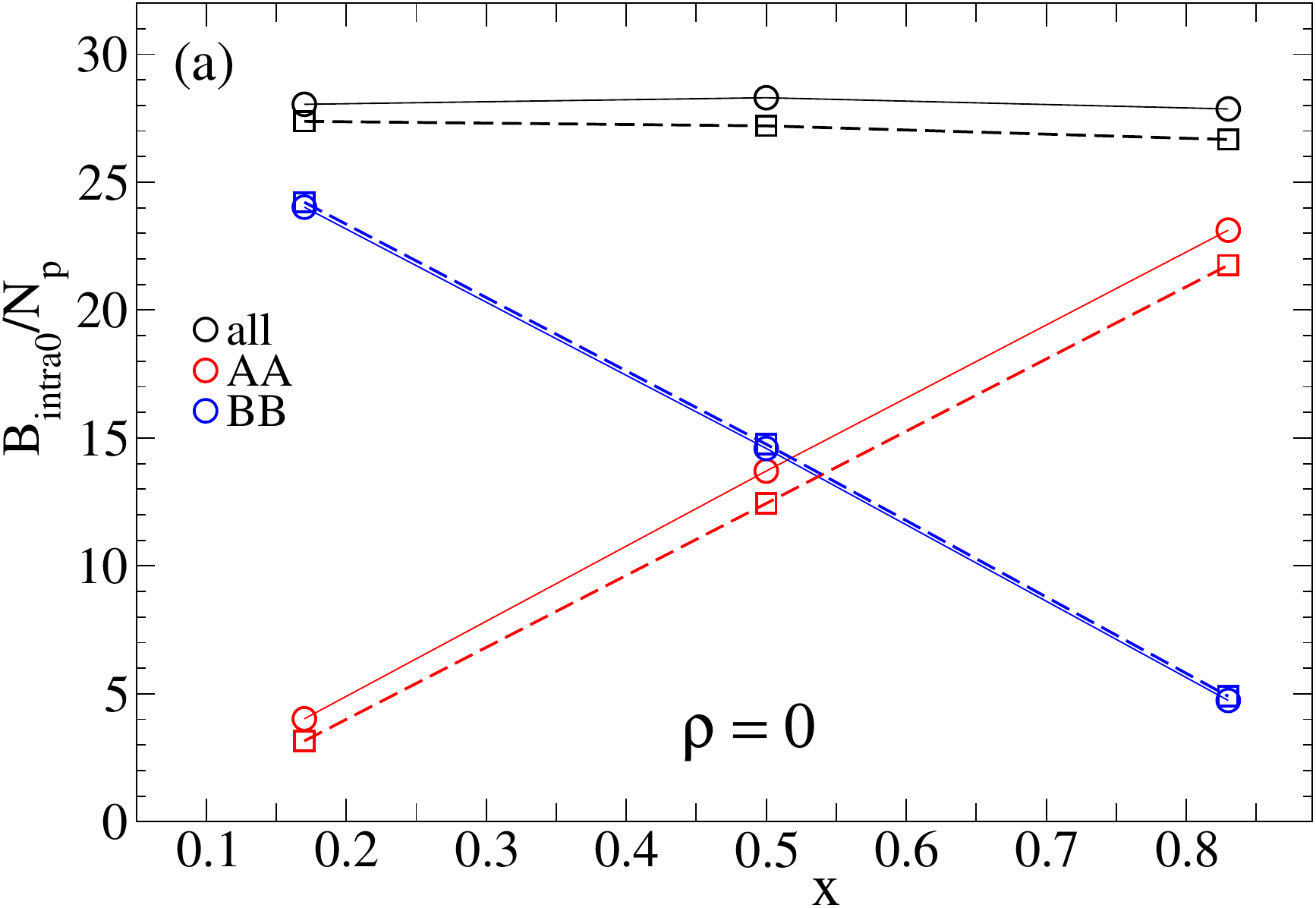}
		\\[5mm]
		\includegraphics[width=0.45\textwidth]{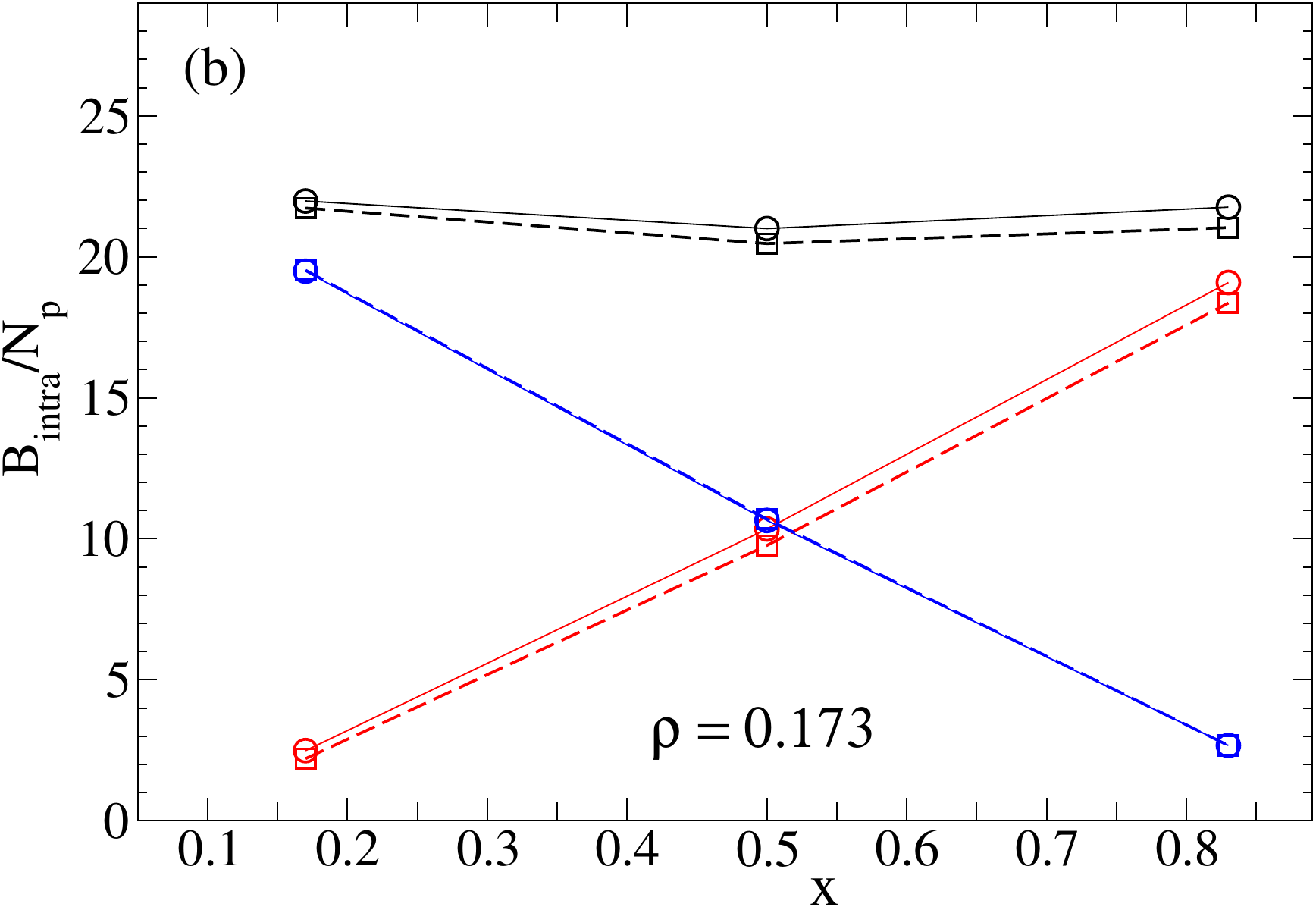}
		\\[5mm]
		\includegraphics[width=0.45\textwidth]{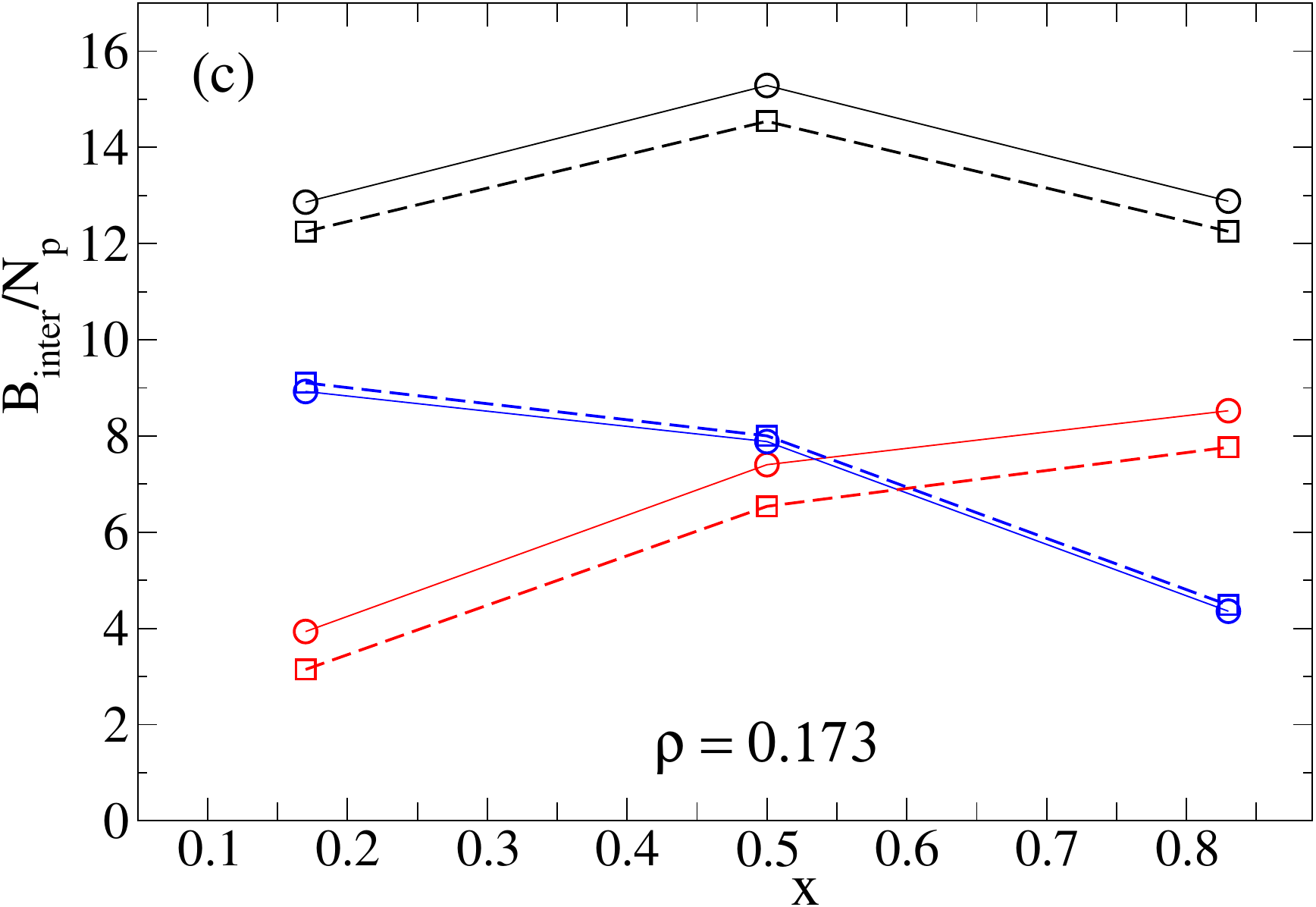}
		\\[2mm]
        \caption{(a) Number of total \textit{i.e.} intra-molecular bonds per chain at infinite  
        dilution. Similarly, (b) and (c) are the intra and inter-molecular bonds per chain at the finite concentration $\rho=0.173$. The solid and dashed lines correspond to 
        ($\epsilon_{\rm b, AA}, \epsilon_{\rm b, BB}$) = (12-15) and (10-17) cases, respectively.}
	\label{nbonds}
\end{figure}
\begin{figure}[h!]
    \centering
    \includegraphics[width=0.5\textwidth]{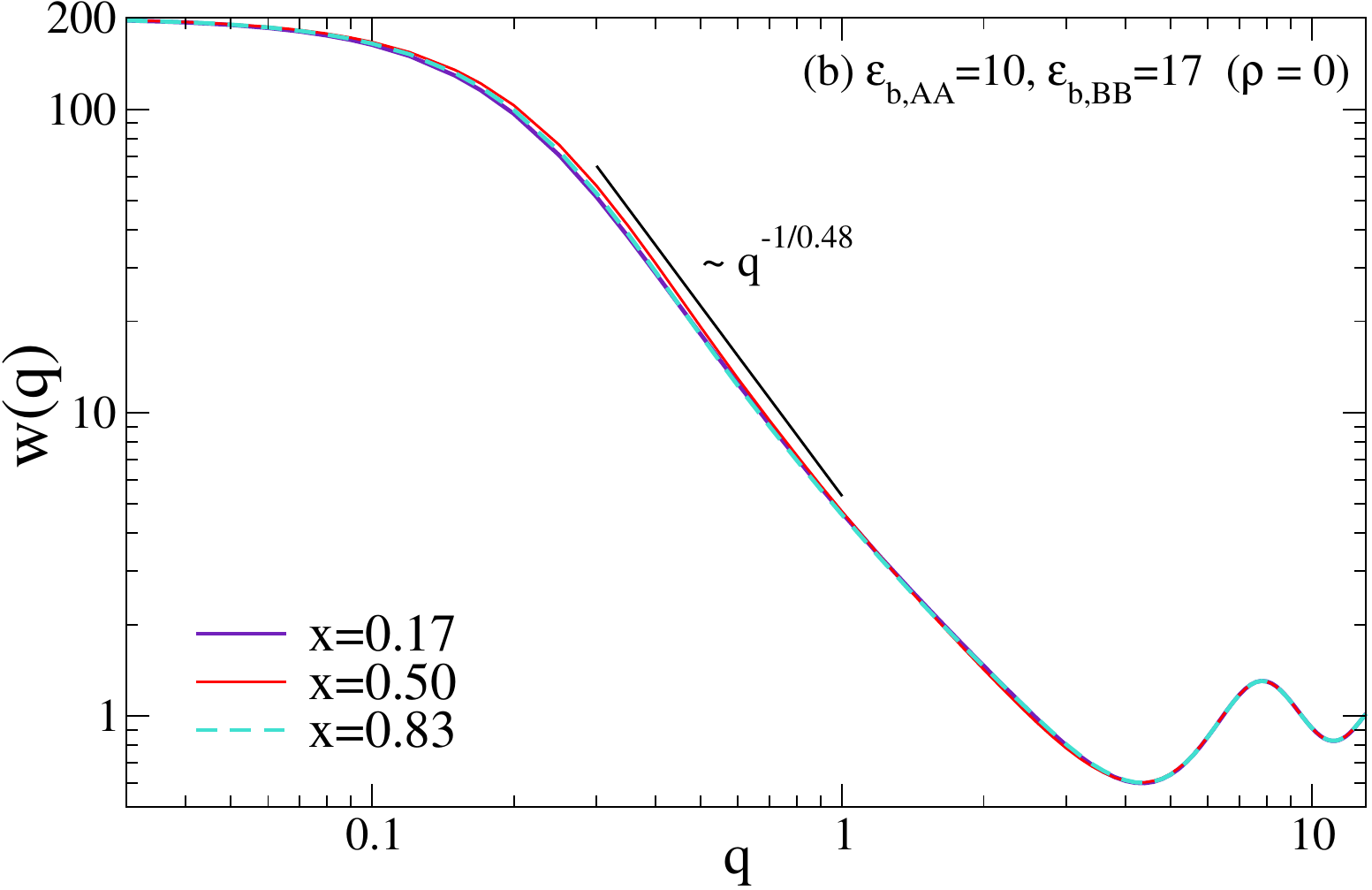}
    \\ [5mm]
    \includegraphics[width=0.5\textwidth]{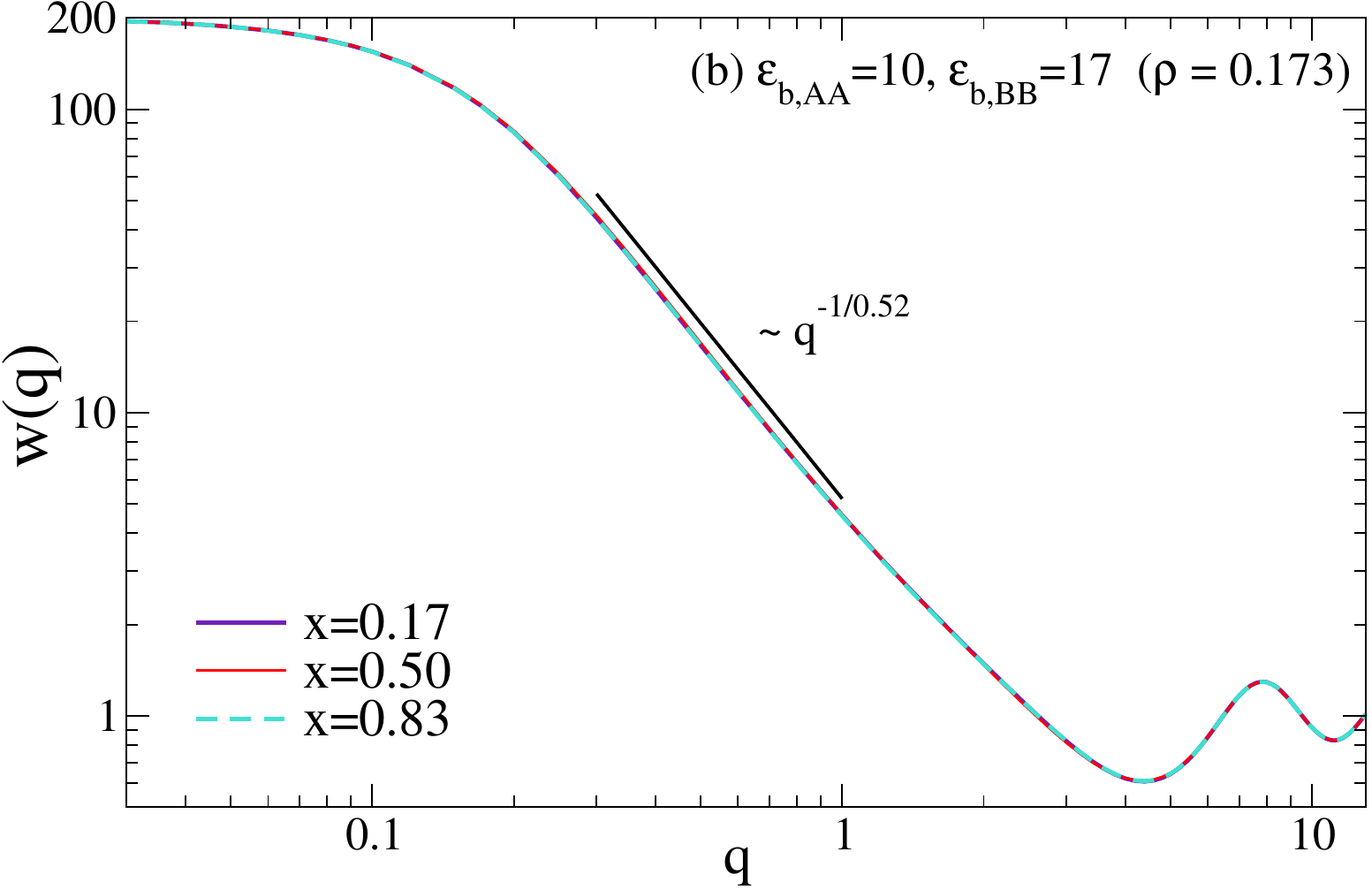}
    \caption{Form factor profiles of the polymers for ($\epsilon_{\rm b,AA}, \epsilon_{\rm b,BB}$) = (10,17) at (a) infinite dilution, and (b) $\rho=0.173$.}
    \label{fig:wq10-17}
\end{figure}
\begin{figure}[h!]
    \centering
    \includegraphics[width=0.5\textwidth]{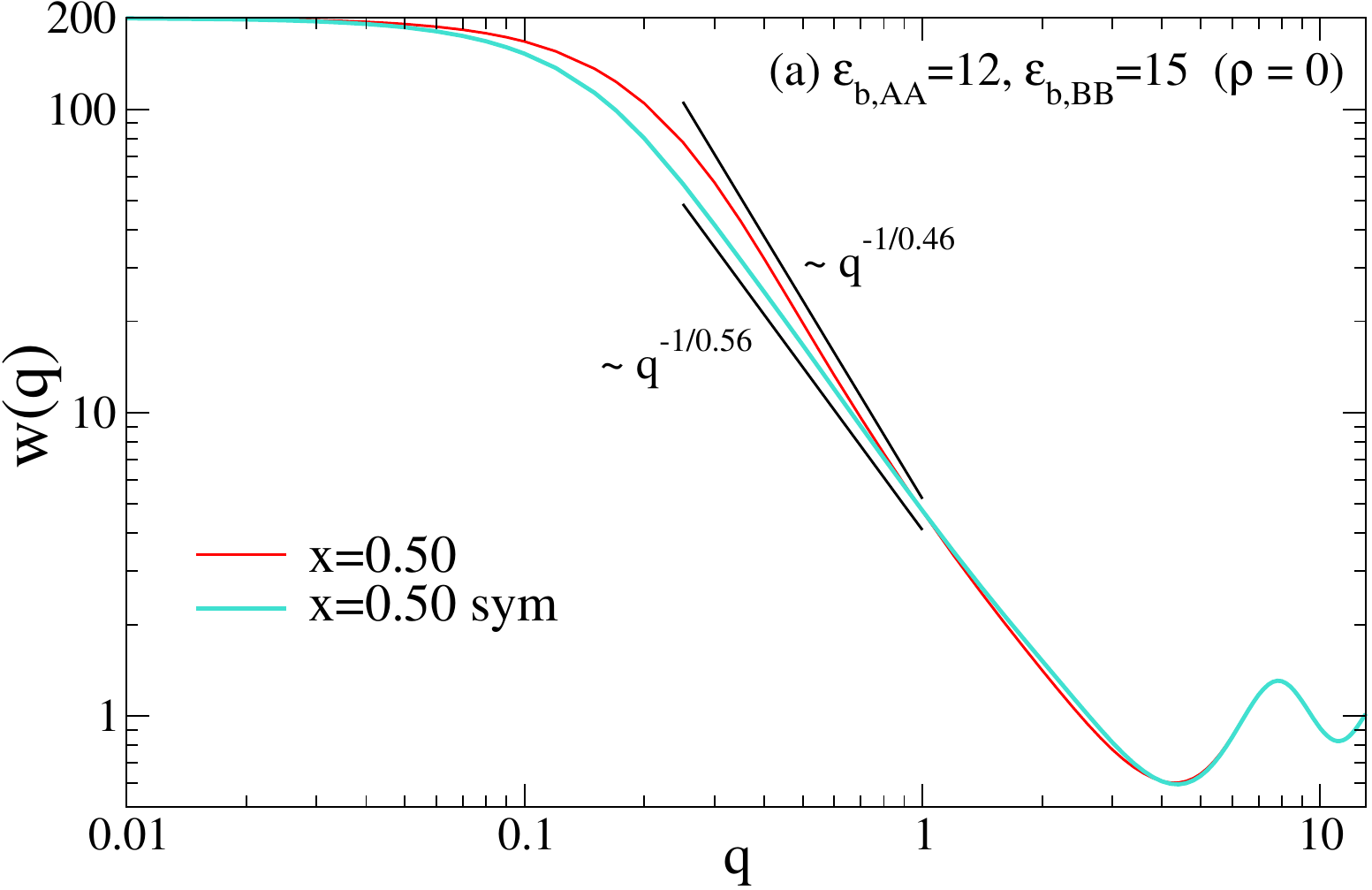}
    \\ [5mm]
    \includegraphics[width=0.5\textwidth]{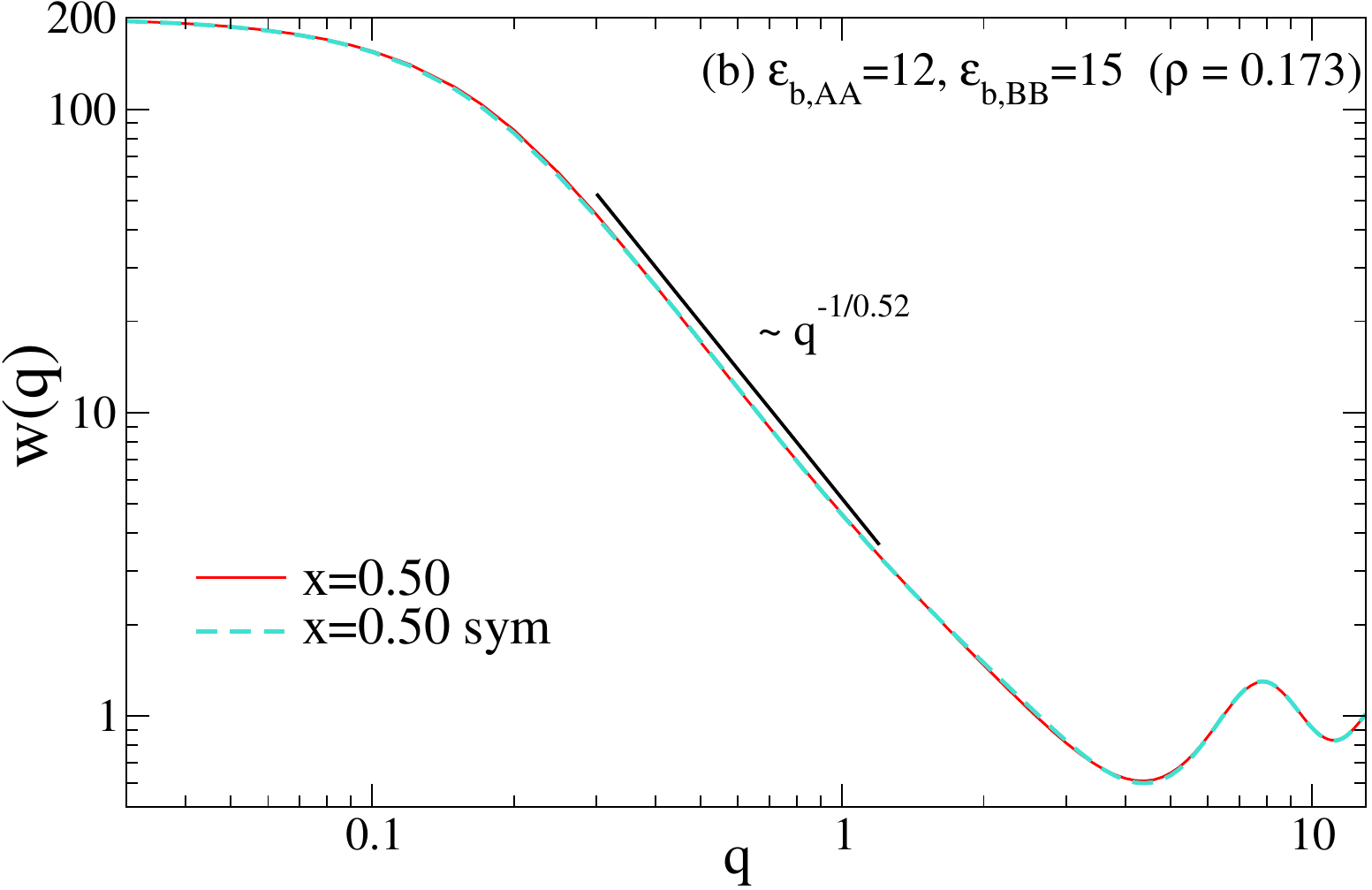}
    \caption{Form factor profiles of the chains for random and symmetric monomer sequences corresponding to ($\epsilon_{\rm b,AA}, \epsilon_{\rm b,BB}$) = (12,15) at (a) infinite dilution,  and (b) $\rho=0.173$.}
    \label{fig:wqrandsymm}
\end{figure}

\noindent\textbf{\large{2. Form factor profiles}}
\newline
Below we present the form factor profiles for ($\epsilon_{\rm b, AA}, \epsilon_{\rm b, BB}$) = (10,17)
case as a function of $x$ at both the densities studied (Fig. \ref{fig:wq10-17}). Similarly, we make
a comparison of the form factor profiles between the symmetric and random cases at $\rho = 0$ and 
0.173 in Fig. \ref{fig:wqrandsymm}. 

\begin{figure}
    \centering
    \includegraphics[width=0.5\textwidth]{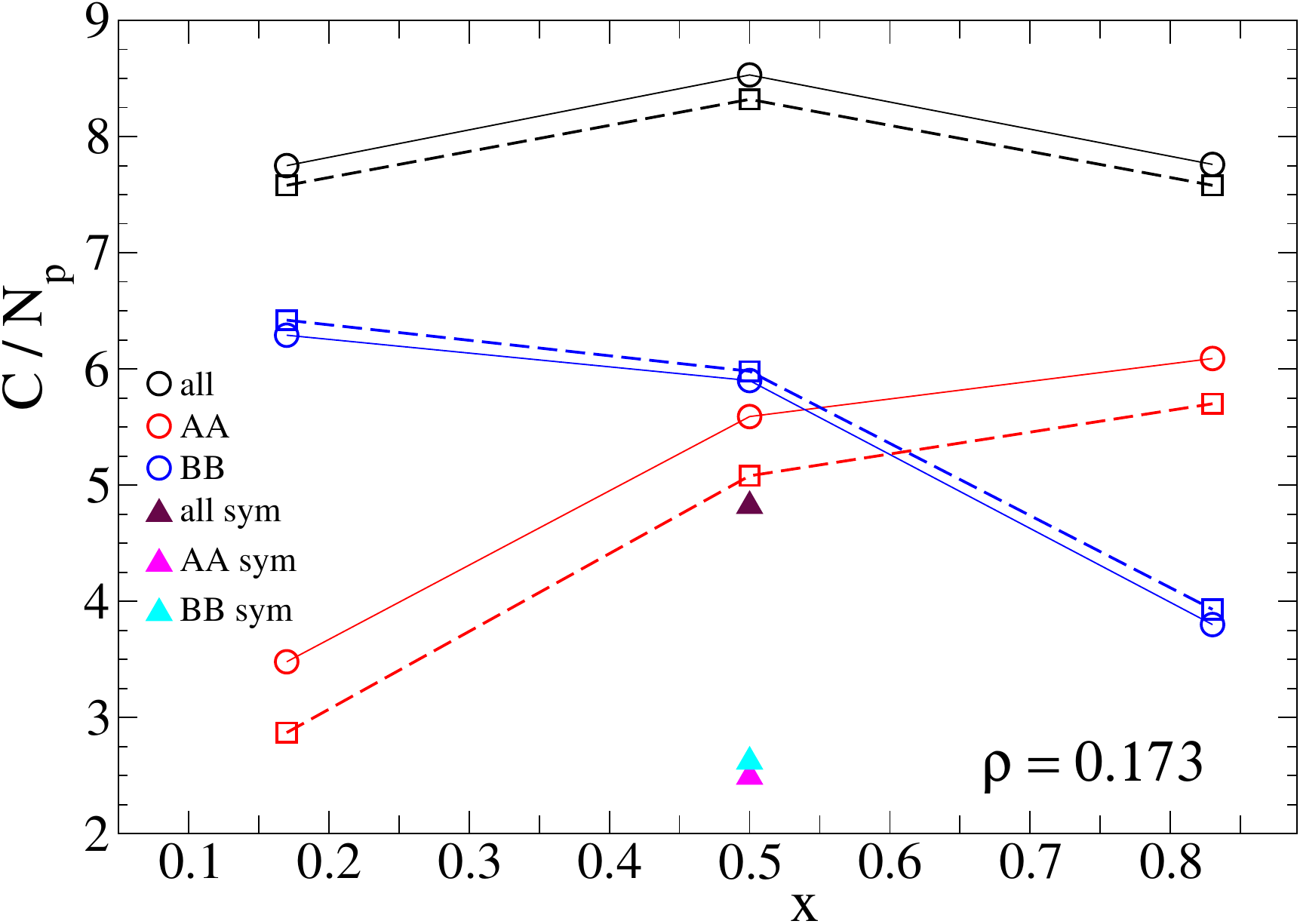}
    \caption{Connectivity per chain for various cases studied at $\rho=0.173$. The solid and dashed lines represent
    ($\epsilon_{\rm b,AA}, \epsilon_{\rm b,BB}$) = (12,15) and (10,17) cases respectively.}
    \label{fig:connectivityperchain}
\end{figure}

\noindent\textbf{\large{3. Connectivity per chain}}
\newline
We define the connectivity ($C$) per chain as the number of different chains a given chain is connected
to. Shown in Fig. \ref{fig:connectivityperchain} are the $C/N_p$ values for all the cases studied at 
$\rho=0.173$. 

\end{document}